\documentclass[a4paper,11pt]{article}
\usepackage{pos}
\usepackage{subfigure} 
\usepackage{setspace}

\newcommand{\mpi}{M_\pi}

\newcommand{\MSb}{\overline{\text{MS}}}
\DeclareMathOperator{\Tr}{Tr}

\begin{document}

\title{Update on Flavor Diagonal Nucleon Charges}

\author*[a]{Sungwoo Park}
\author[b]{Tanmoy Bhattacharya}
\author[b]{Rajan Gupta}
\author[c]{Huey-Wen Lin}
\author[b,c]{Santanu Mondal}
\author[d]{Boram Yoon}
\affiliation[a]{Thomas Jefferson National Accelerator Facility,
  12000 Jefferson Avenue, Newport News, VA 23606, USA}
\affiliation[b]{Theoretical Division T-2, Los Alamos National Laboratory, Los Alamos, NM 87545, USA}
\affiliation[c]{Department of Physics and Astronomy, Michigan State University, East Lansing, MI 48824, USA}
\affiliation[d]{Computer, Computational and Statistical Science Division CCS-7, Los Alamos National Laboratory, Los Alamos, NM 87545, USA}

\emailAdd{sungwoo@jlab.org}
\emailAdd{tanmoy@lanl.gov}
\emailAdd{rajan@lanl.gov}
\emailAdd{hueywen@msu.edu}
\emailAdd{santanu.sinp@gmail.com}
\emailAdd{byoon@nvidia.com}

\abstract{This talk provides an update on the calculation of matrix
  elements of flavor diagonal axial, scalar and tensor quark bilinear
  operators between the nucleon ground state. The simulations are done
  using Wilson-clover fermions on a sea of eight 2+1+1-flavor HISQ
  ensembles generated by the MILC collaboration. We discuss the signal
  in the sum of the connected and disconnected contributions for the up, down and strange quarks, 
  control over fits to remove excited state contamination, 
  and the simultaneous chiral-continuum fit
  used to extract the charges. }

\FullConference{%
  The 39th International Symposium on Lattice Field Theory (Lattice2022),\\
  8-13 August, 2022 \\
  Bonn, Germany 
}

\maketitle


\section{Introduction}
\label{sec:intro}

This talk updates the calculation of the matrix elements of flavor 
diagonal axial, scalar and tensor quark bilinear operators between the nucleon ground state from which we extract the charges $g_{A,S,T}^{u,d,s}$. The motivation for these calculations 
and much of the methodology used have already been 
published: for $g_{A,T}^q$ in
Refs.~\cite{Gupta:2018qil,Gupta:2018lvp,Lin:2018obj}, and for $g_S^q$
in Ref.~\cite{Gupta:2021ahb}.  
Here we will focus on describing the progress since Lattice 2021~\cite{Park:2021ggz}. Since the final analysis is 
still ongoing, all the results presented here should be considered 
preliminary unless otherwise stated.

\begin{figure}[ht] 
  \centering
  \begin{subfigure}[Nucleon charges]{
      \includegraphics[width=0.235\linewidth]{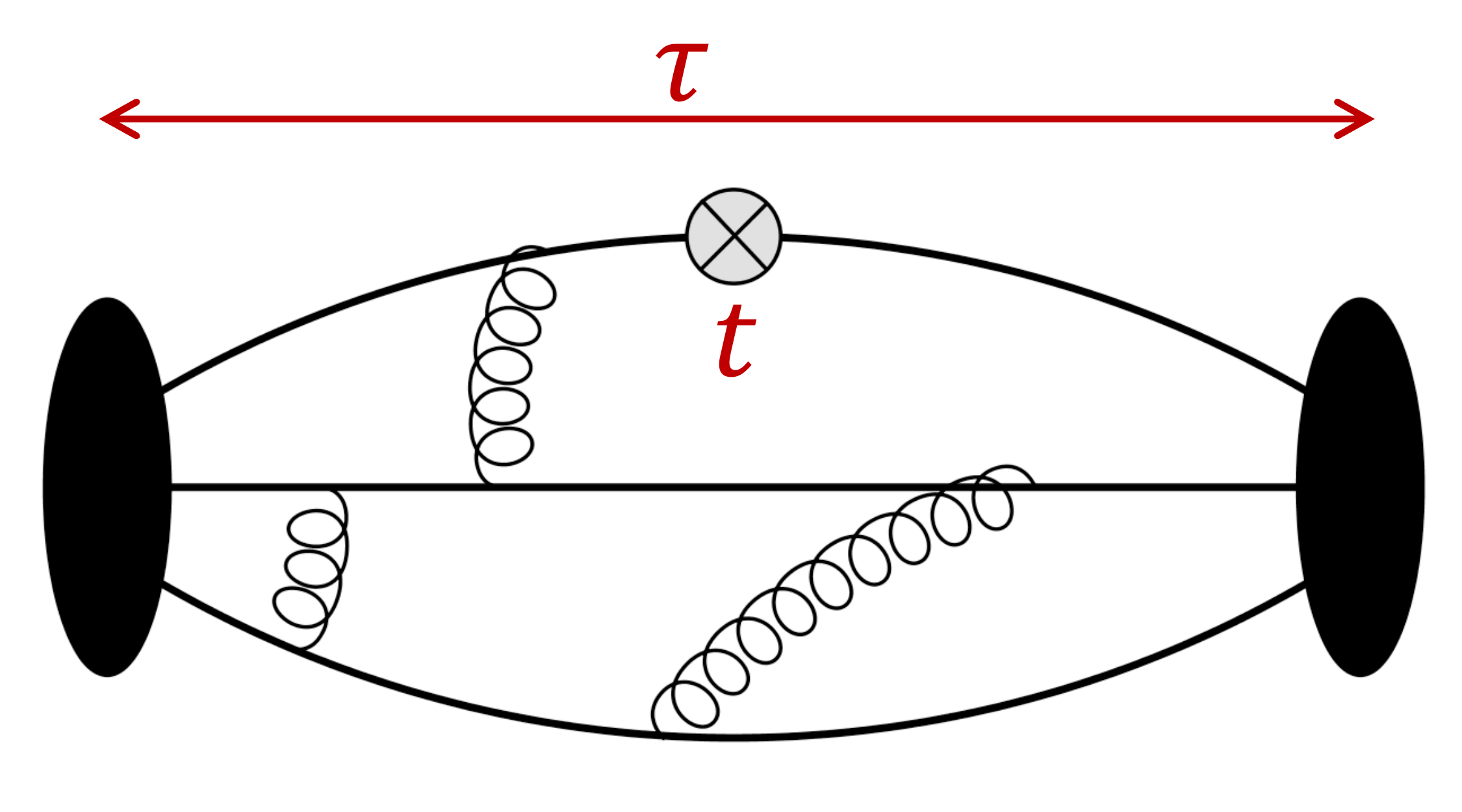}
      \includegraphics[width=0.235\linewidth]{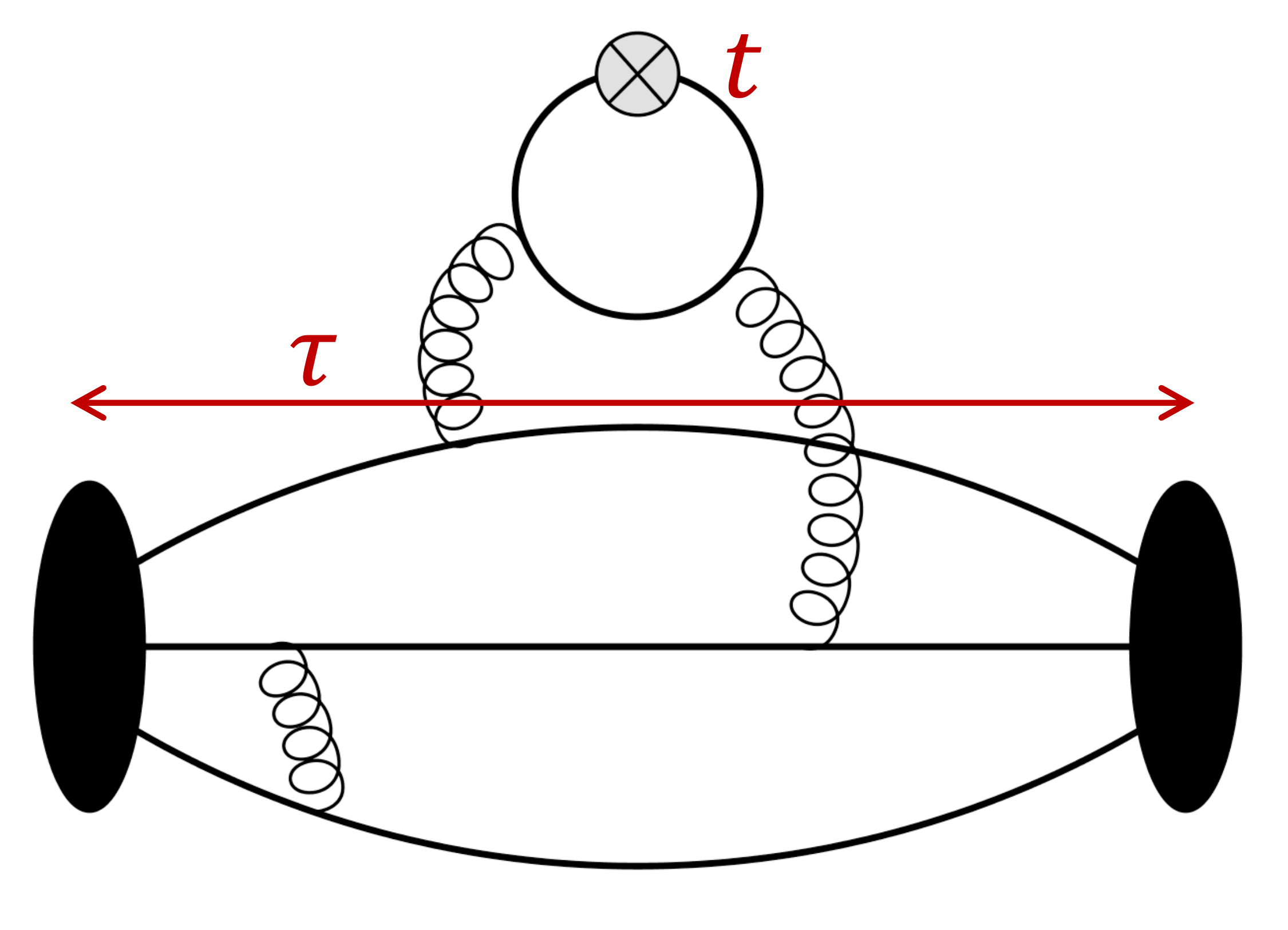}
      \label{fig:diag_g}
    }
  \end{subfigure}
  \begin{subfigure}[NPR]{
      \includegraphics[width=0.235\linewidth]{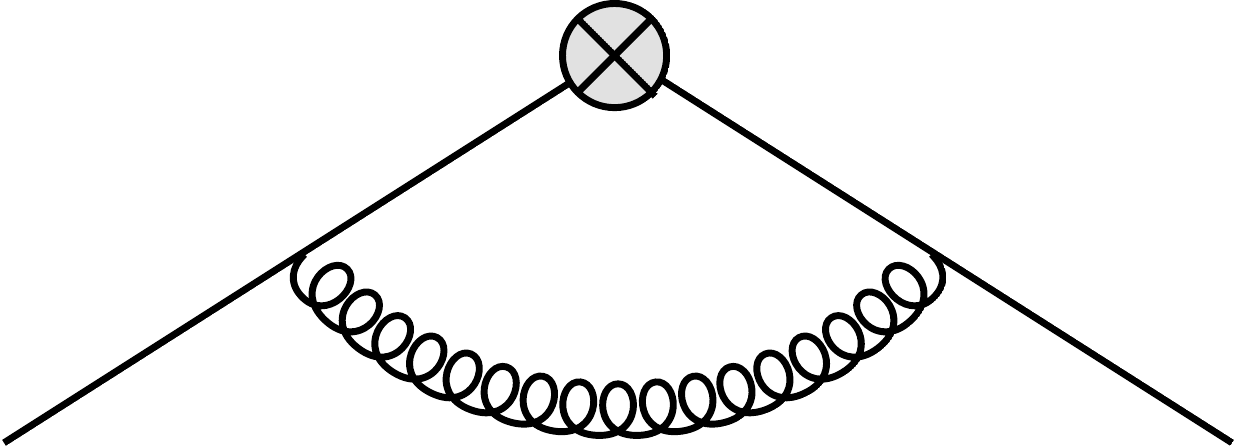}
      \includegraphics[width=0.235\linewidth]{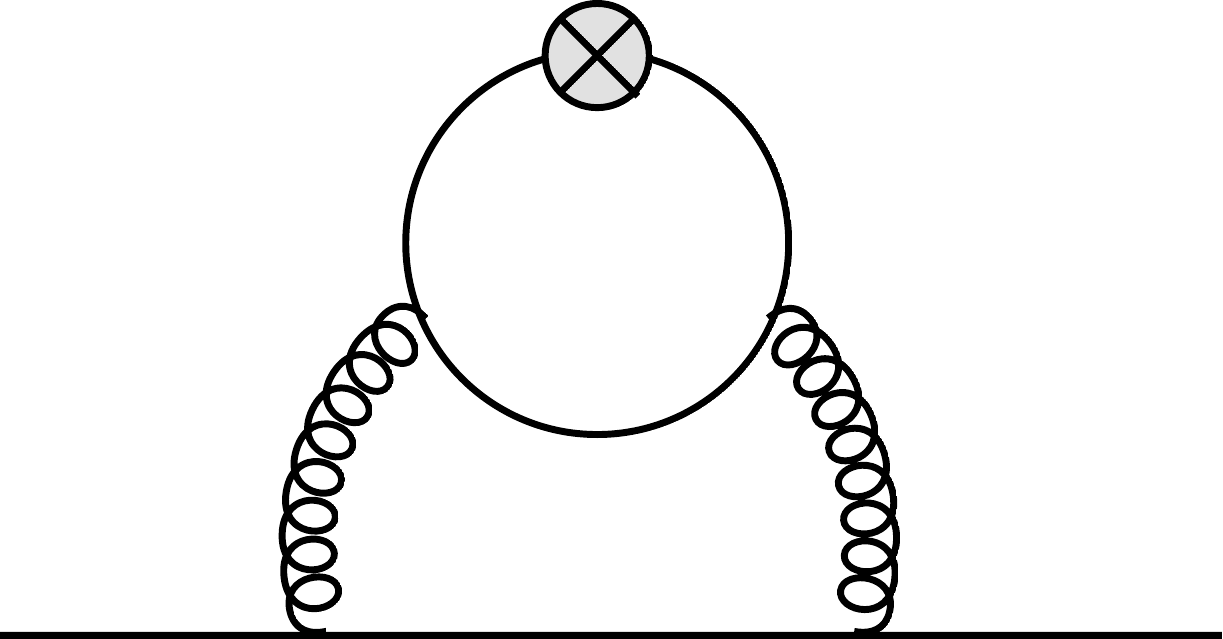}
      \label{fig:diag_npr}
    }
  \end{subfigure}
  \caption{The connected and disconnected diagrams calculated (i) for flavor diagonal nucleon charges,  and (ii) 
    non-perturbative renormalization in the RI-sMOM scheme using quark states in Landau gauge.} 
\end{figure}

The calculations  have been done on 
eight 2+1+1-flavor HISQ ensembles generated by the MILC
collaboration~\cite{Bazavov:2012xda} with correlation functions 
calculated using Wilson-clover valence fermions with the light ($m_u = m_d$ in the 
isospin symmetric theory) and strange quark masses 
tuned to reproduce the sea (HISQ) $M_\pi$ and
$M_{s \bar s}$ values. The parameters of these ensembles are given in
Table~\ref{tab:hisq}. This set includes one physical $M_\pi\approx
138$ MeV ensemble (labeled as $a09m130$) at $a\approx 0.09$ fm and
$M_\pi L\approx 3.9$. In addition to the nucleon 2-point
functions~\cite{Gupta:2018qil}, we calculate the quark-line diagrams for the connected~\cite{Gupta:2018qil} and
disconnected~\cite{Gupta:2018lvp,Lin:2018obj,Gupta:2021ahb} contributions to 3-point
functions illustrated in the two left panels in Fig.~\ref{fig:diag_g},
and the analogous quark level diagrams in Landau gauge for calculating
the renormalization constants in the RI-sMOM scheme in Fig.~\ref{fig:diag_npr}.

\begin{table}[hb] 
  \vspace{-0.5mm}
\center  
\resizebox{\textwidth}{!}{
\begin{tabular}{l|cccc|ccccccccccccccccc}
Ensemble ID    & $a$ (fm) & $M_\pi$ (MeV)    & $M_\pi L$  &  $L^3\times T$ & $N_\text{conf}^l$ & $N_\text{src}^l$ & $N_\text{conf}^s$ &
    $N_\text{src}^s$ & $N_\text{LP}/N_\text{HP}$\\\hline
$a15m310$    & 0.1510(20) & 320(5)      & 3.93    & $16^3\times 48$   & 1917 & 2000  & 1917 & 2000  & 50    \\
$a12m310$    & 0.1207(11) & 310(3)      & 4.55    & $24^3\times 64$   & 1013 & 10000 & 1013 & 8000  & 50    \\
$a12m220$    & 0.1184(10) & 228(2)      & 4.38    & $32^3\times 64$   & 958  & 11000 & 870  & 5000  & 30--50\\
$a09m310$    & 0.0888(8) &  313(3)      & 4.51    & $32^3\times 96$   & 1017 & 10000 & 1024(*)  & 6000  & 50    \\
$a09m220$    & 0.0872(7) &  226(2)      & 4.79    & $48^3\times 96$   & 712  & 8000  & 847  & 10000 & 30--50\\
$a09m130$    & 0.0871(6) &  138(1)      & 3.90    & $64^3\times 96$   & 1270 & 10000 & 541+453(*)  & 10000+4000 & 50    \\
$a06m310$    & 0.0582(4) &  320(2)      & 3.90    & $48^3\times 144$  & 808  & 12000 & 948+28(*)  & 10000+4000 & 50    \\
$a06m220$    & 0.0578(4) &  235(2)      & 4.41    & $64^3\times 144$  & 1001 & 10000 & 1002 & 10000 & 50    \\
\end{tabular}}
\vspace{-2mm}
\caption{The ensembles and the statistics used for the  calculation of disconnected contributions, including updates to Refs.~\cite{Lin:2018obj,Gupta:2018lvp,Park:2021ggz}. Statistics for the connected contributions are the same as in
  Ref.~\cite{Gupta:2018qil}.  $N_\text{conf}^{l,s}$ is the number of gauge configurations analyzed for light ($l$) and strange ($s$) flavors. $N_\text{src}^{l,s}$ is the number of random sources used
  per configurations, and $N_\text{LP}/N_\text{HP}$ is the ratio of
  low/high precision measurements.  The $N_{\rm conf}^s$ marked with (*) have been updated since 2021 \cite{Park:2021ggz}. }
\label{tab:hisq}
\end{table}
\vspace{-0.9mm}


\section{Details of 2-point and 3-point function analysis}
\label{sec:details}

The parameters of quark propagators calculated on HYP smeared
lattices using Wuppertal smearing are given in
Ref.~\cite{Gupta:2018qil}.  To construct all the 2- and 3-point correlation 
functions, the nucleon interpolating operator 
$\mathcal{N}(x) = \epsilon^{abc} \left[ {q_1^a}^T(x) C \gamma_5 \frac{(1 \pm
    \gamma_4)}{2} q_2^b(x) \right] q_1^c(x) $ is used
both at the source and the sink.  
To extract the charges $g_{A,S,T}^{u,d,s}$ from forward matrix elements, 
all 3-point functions are calculated with zero momentum  
projection of both the nucleon state at the sink and  the operator insertion: 
$ C^{\text{3pt}}_\Gamma(t;\tau)=\Tr [ {\cal P} \langle 0|\mathcal{N}(\tau)   O_\Gamma(t,{\textbf q=0}) \bar{\mathcal{N}}(0,{\textbf p=0})|0 \rangle ]$ with 
$O_\Gamma^q= \bar q \Gamma q$, $q \in \{u,d,s\}$, and
$\cal P$ is the spin
projection defined in Ref.~\cite{Bhattacharya:2015wna} for the various cases. 

Flavor diagonal 3pt functions are the sum of connected (conn) and disconnected (disc)
contributions illustrated in Fig.~\ref{fig:diag_g}: 
$C^{\text{3pt}}_\Gamma(t;\tau) \equiv C^{\text{conn}}_\Gamma(t;\tau) + C^{\text{disc}}_\Gamma(t;\tau)$. For the scalar case, the disconnected contribution is calculated using
the vacuum subtracted operator $O_S^q - \langle O_S^q \rangle$.  The
calculation of the quark loop with zero-momentum operator insertion is
estimated stocastically using $Z_4$ random noise sources as explained in
Ref.~\cite{Bhattacharya:2015wna}. Note that in our previous works~\cite{Lin:2018obj,Gupta:2018lvp}, the
fits to remove ESC in $C^{\text{conn}}_\Gamma(t;\tau)$ and
$C^{\text{disc}}_\Gamma(t;\tau)$ were done separately, as was the
chiral-continuum (CC) extrapolations of $g_\Gamma^{q,\text{disc}}$ and
$g_\Gamma^{q,\text{conn}}$. This introduced an unquantified
systematic~\cite{Lin:2018obj} that has now been removed by making a
simultaneous fit to $C^{\text{3pt}}_\Gamma(t;\tau)$ and $C^\text{2pt}(\tau)$ and
extrapolating $g_\Gamma^{q}$. 

The bare charges, $g_\Gamma^{q;\text{bare}}$, are obtained  from the
ground state matrix elements $\langle 0 | O_\Gamma^q | 0 \rangle$ 
extracted from fits to  the spectral decomposition of the spin projected $C^{\textrm{3pt}}_\Gamma (t;\tau)$:
\begin{align}
    C^{\textrm{3pt}}_\Gamma (t;\tau) &=\sum_{i,j=0} \mathcal{A}_i
    \mathcal{A}_j^\ast \langle i |O_\Gamma^q| j \rangle e^{-M_i t - M_j (t- \tau)}
    \quad {\rm with} \quad \langle 0 | O_\Gamma^q | 0 \rangle = g_{\Gamma}^q \,. 
\label{eq:3pt-sd}
\end{align}
The challenge to extracting  
$\langle 0 | O_\Gamma | 0 \rangle$ from fits to $C^{\textrm{3pt}}_\Gamma$
is removing excited state contributions (ESC) which are
observed to be large at source-sink separation $\tau\approx 1.5$ fm 
beyond which the signal degrades due to
the $e^{(M_N-3/2M_\pi)\tau}$ increase in noise.  With the current
statistics, we are only able to keep one excited state in Eq.~\eqref{eq:3pt-sd}, 
and fits leaving $M_1$ a free parameter
are not stable in many cases.\looseness-1

The nucleon spectrum $M_i$ and amplitude $\mathcal{A}_0$ needed to analyze $ C^{\text{3pt}}_\Gamma(t;\tau)$ are 
obtained from the spectral decomposition of the 2pt function, $C^\textrm{2pt}(\tau)=\sum_{i=0} | \mathcal{A}_i|^2 e^{-M_i \tau}$, truncated at four states. 
We carry out two types of analyses: (i) The ``standard''  
fit to $C^\textrm{2pt}(\tau)$ uses wide
priors for all the excited-state amplitudes, $\mathcal{A}_i$, and
masses, $M_i$, i.e., the priors are only used to stabilize the
fits. In these fits, $M_1 \gtrsim 1.5$~GeV. (ii) The ``$N\pi$'' fit in which a narrow prior is used for
$M_1$ with the central value given by the non-interacting energy of the
lowest allowed $N \pi$ or $N \pi \pi$ state on the lattice. The resulting values of $\mathcal{A}_0$
and the $M_i$ are then used as inputs in the analysis of the 3-point
functions. In practice, we fit $C^\textrm{2pt}(\tau)$ and $C^{\textrm{3pt}}_\Gamma (t;\tau)$ simultaneously. The important point is that the mass gap, $M_1-M_0$, in the two analyses is
significantly different, however, the augmented $\chi^2$ minimized in the fits is
essentially the same. Since the two fit strategies are not distinguished by the 
$\chi^2$, we examine the sensitivity of the results for the charges 
to the two $M_1$ and use the difference to appropriately 
estimate an associated systematic uncertainty. 

Also, the ESC analysis is repeated to
quantify model variation of results by choosing data with different
set of $(\tau,t)$ values and the number of excited states (2- or 3-state) in
the ansatz of Eq.~\eqref{eq:3pt-sd}. The final ESC analysis results
are taken to be the weighted average with the Akaike information
criteria weight $\text{exp}[-(\chi^2-2N_\text{dof})/2]$
\cite{Jay:2020jkz}.

\looseness-1
  
Another challenge to distinguishing between ``standard'' and ``$N\pi$'' analysis
strategies is that the difference in the corresponding $M_1$ becomes
significant only for $M_\pi \lesssim 200$~MeV, which in our setup
means only in the $a091m130$ ensemble. Previous works show that the
difference in axial and tensor charges, $g_{A,T}$, from the two strategies is
small~\cite{Park:2021ypf}.  For the isoscalar scalar charge
$g_S^{u+d}$, $\chi$PT suggests a large contribution from the $N\pi$ and
$N\pi\pi$ states \cite{Gupta:2021ahb}. As explained in
Ref.~\cite{Gupta:2021ahb}, this leads to a large difference in the value of the
pion-nucleon sigma term. The intent of doing the full analysis with both strategies 
is to quantify these differences and understand which states contribute. \looseness-1

The renormalization of the axial, scalar and tensor operators is carried out 
in the 3-flavor theory (we have explicitly evaluated the $3 \times 3$ 
matrices accounting for flavor mixing)
using the RI-sMOM lattice scheme and then converting to $\MSb$ scheme at
2 GeV as described in Ref.~\cite{Park:2021ggz}.

\begin{figure}[p] 
  \center
      \includegraphics[width=0.30\linewidth]{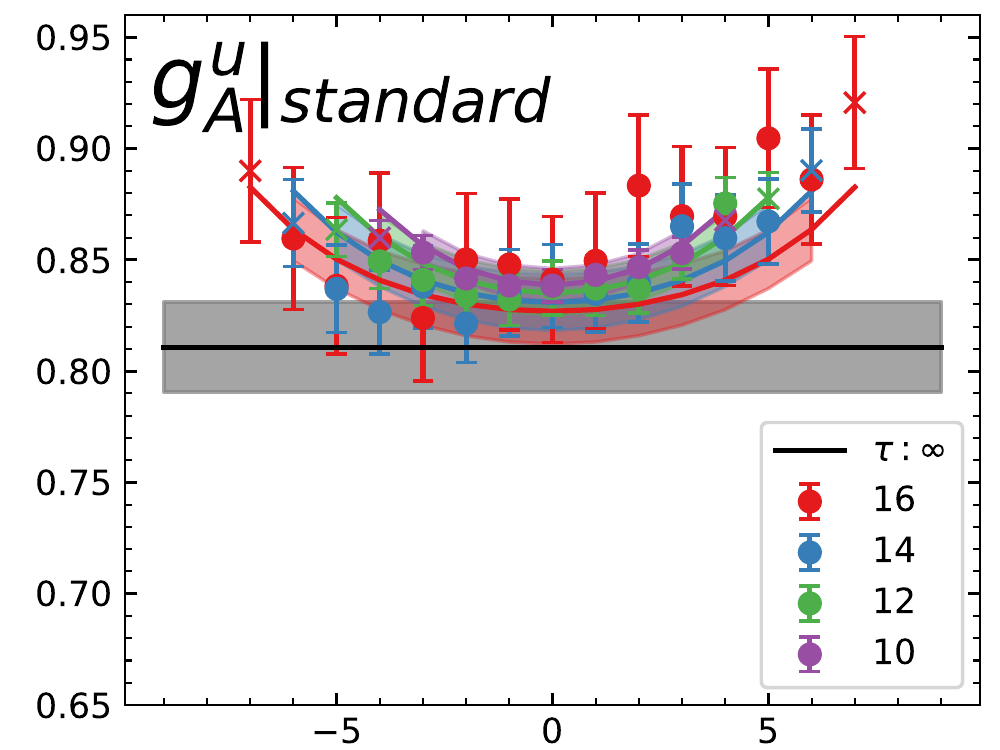}
      \includegraphics[width=0.30\linewidth]{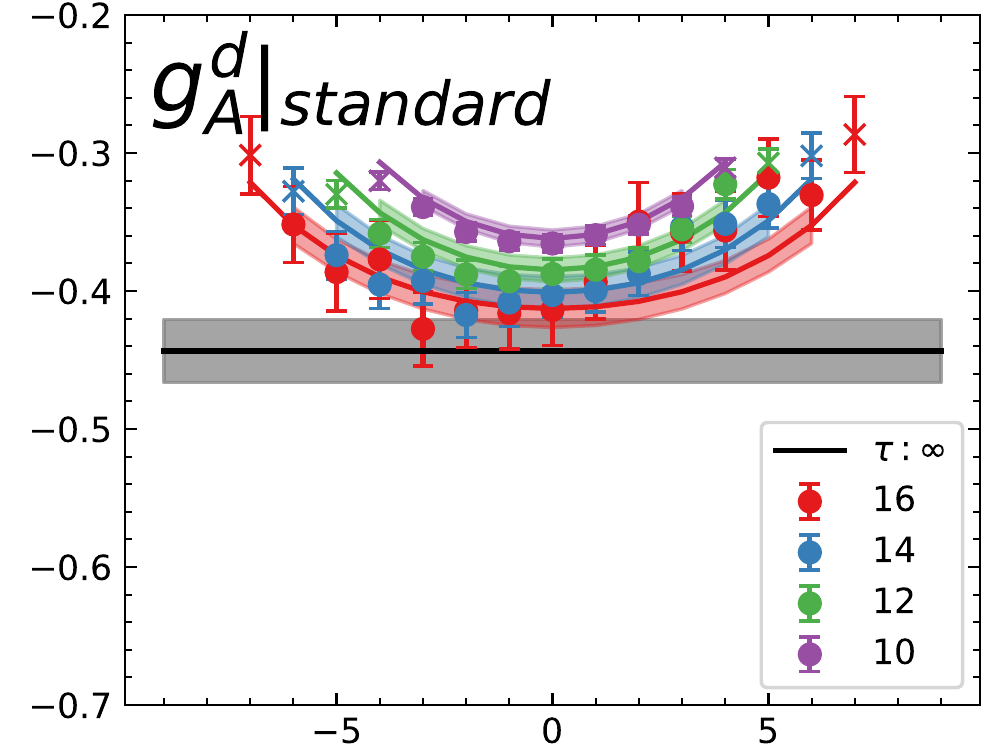}
      \includegraphics[width=0.30\linewidth]{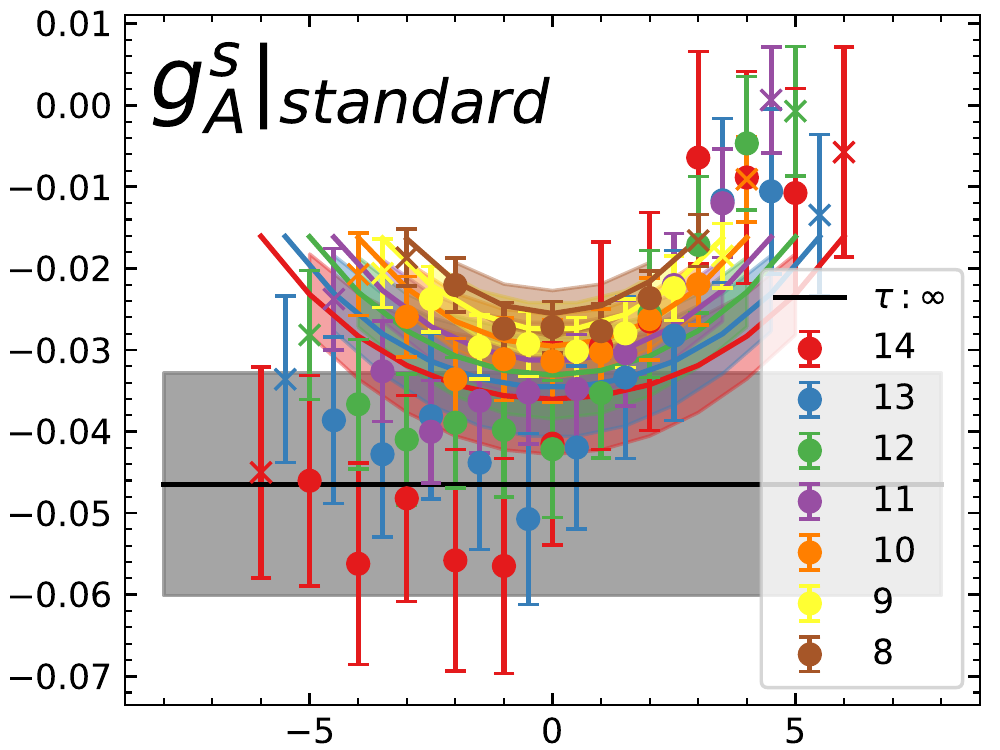}

      \includegraphics[width=0.30\linewidth]{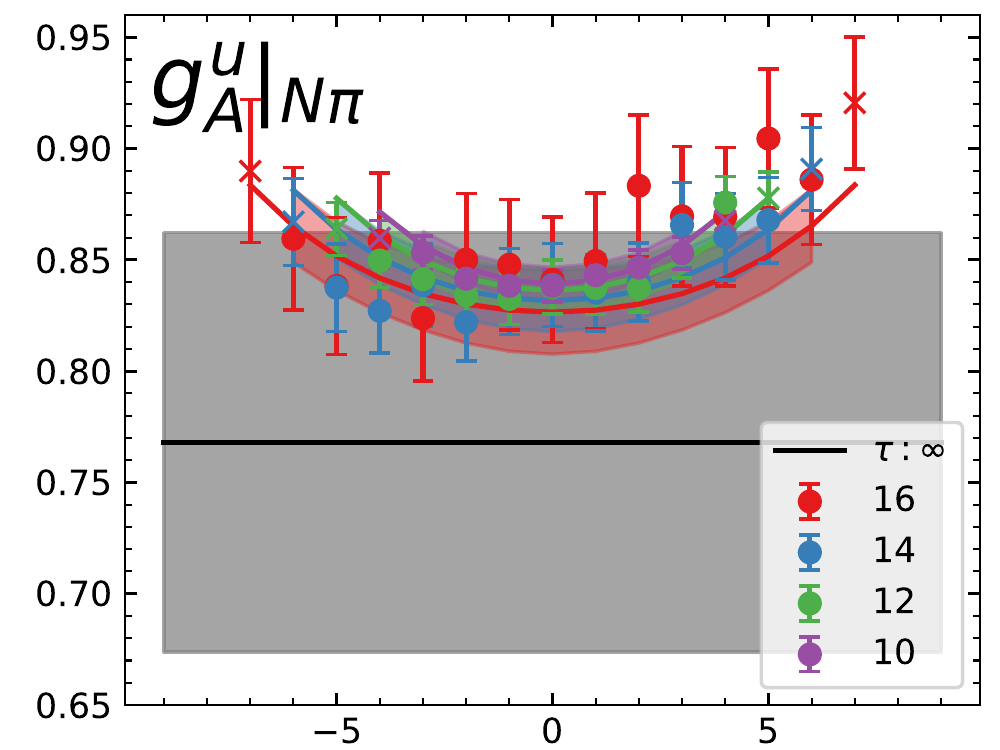}   
      \includegraphics[width=0.30\linewidth]{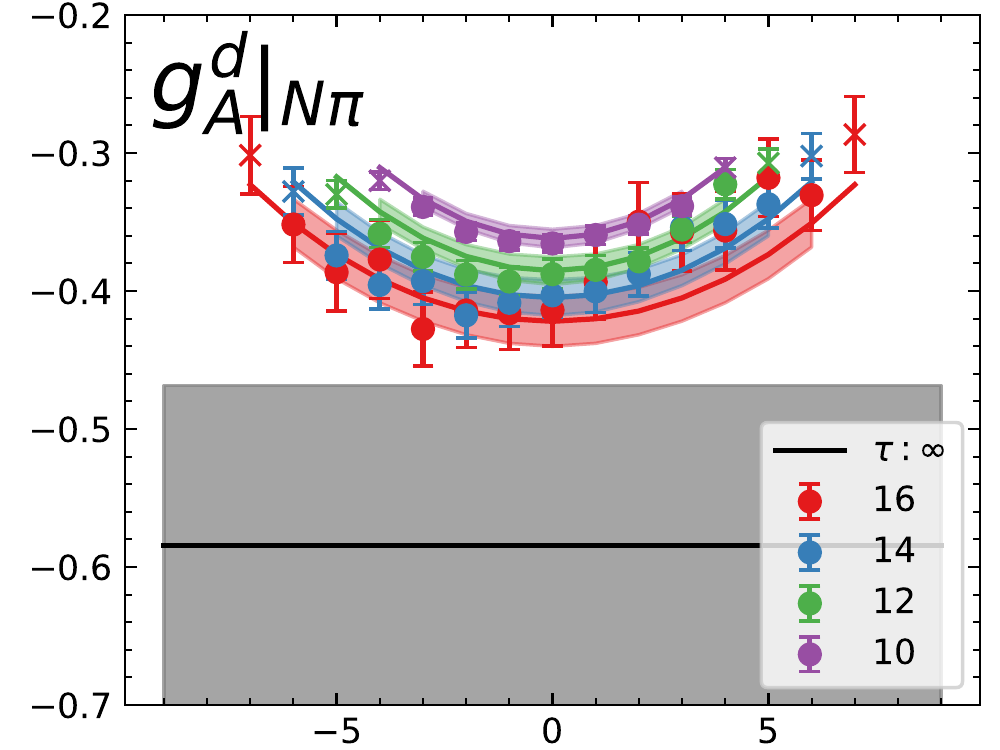}
      \phantom{\includegraphics[width=0.30\linewidth]{figs/3pt_AIC/gAd_4Npi_a09m130.pdf}}

      \includegraphics[width=0.30\linewidth]{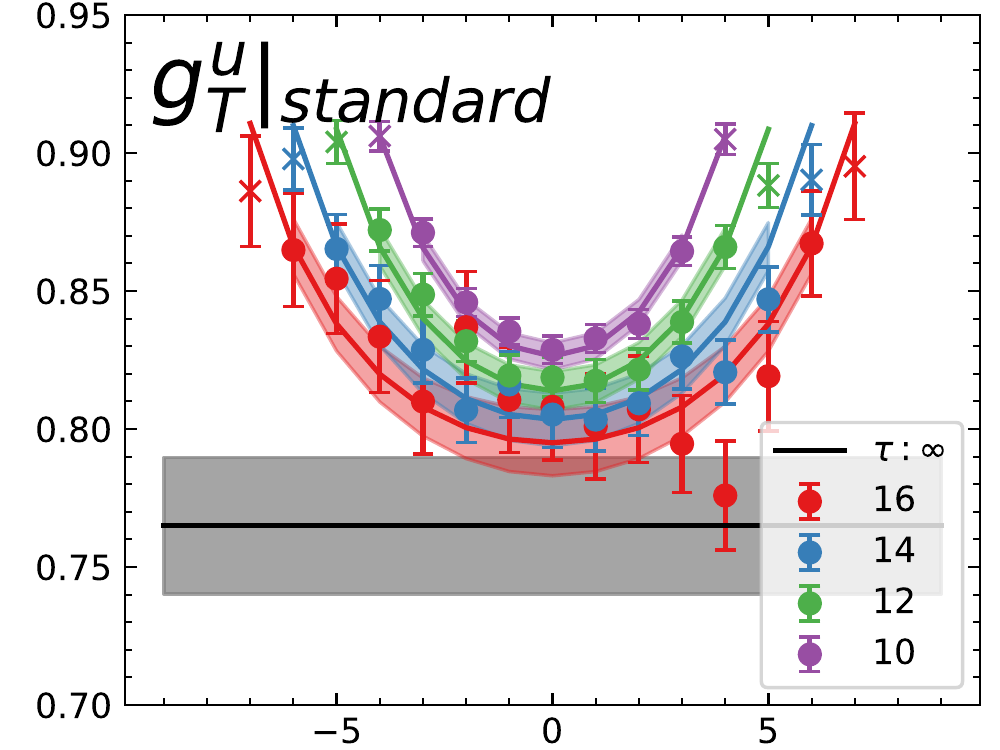}  
      \includegraphics[width=0.30\linewidth]{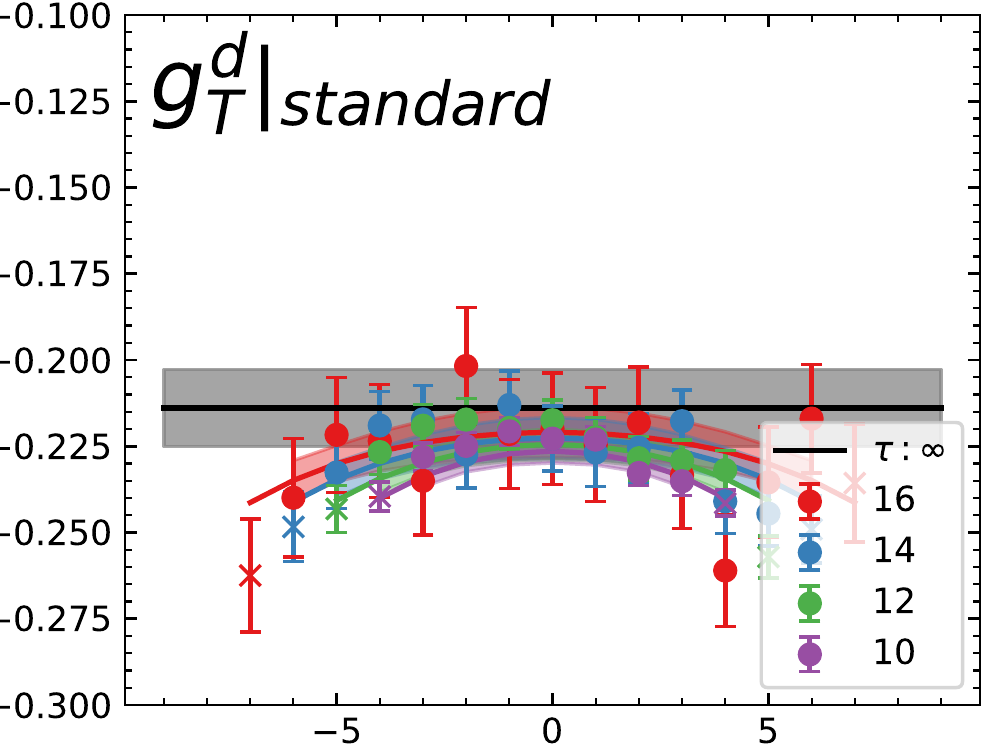}
      \includegraphics[width=0.30\linewidth]{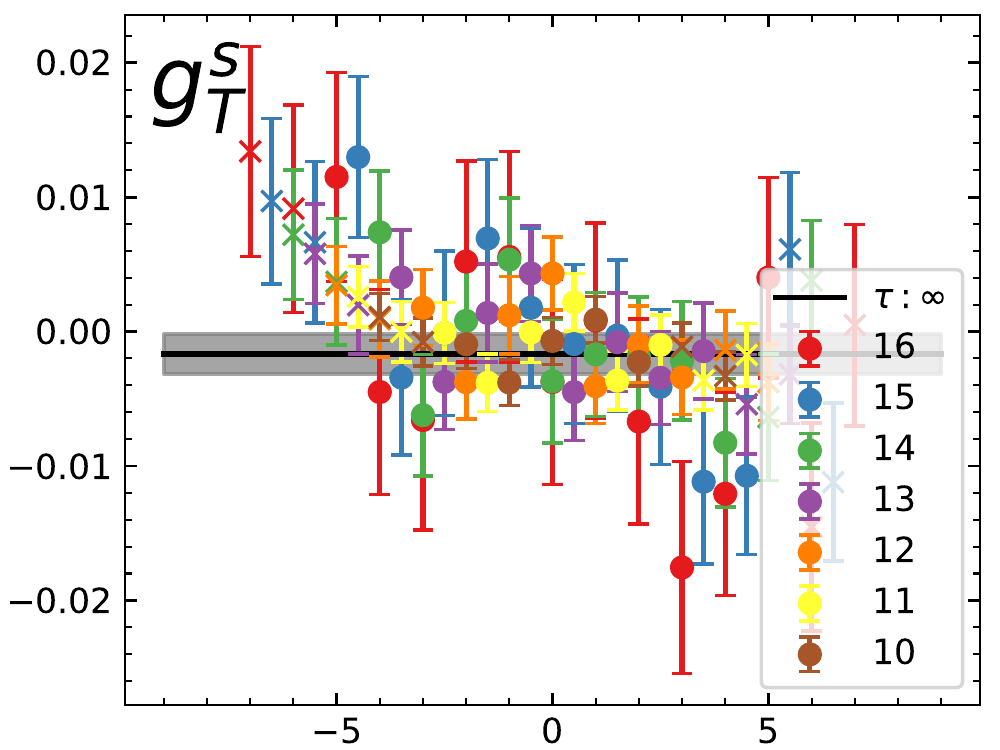}

      \includegraphics[width=0.30\linewidth]{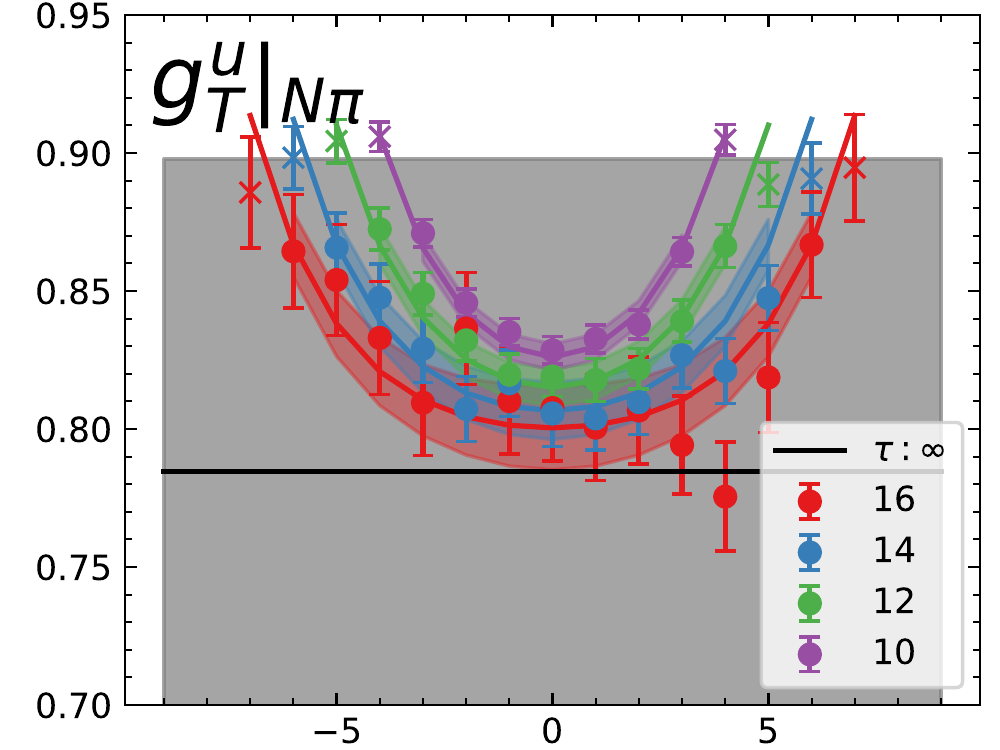}
      \includegraphics[width=0.30\linewidth]{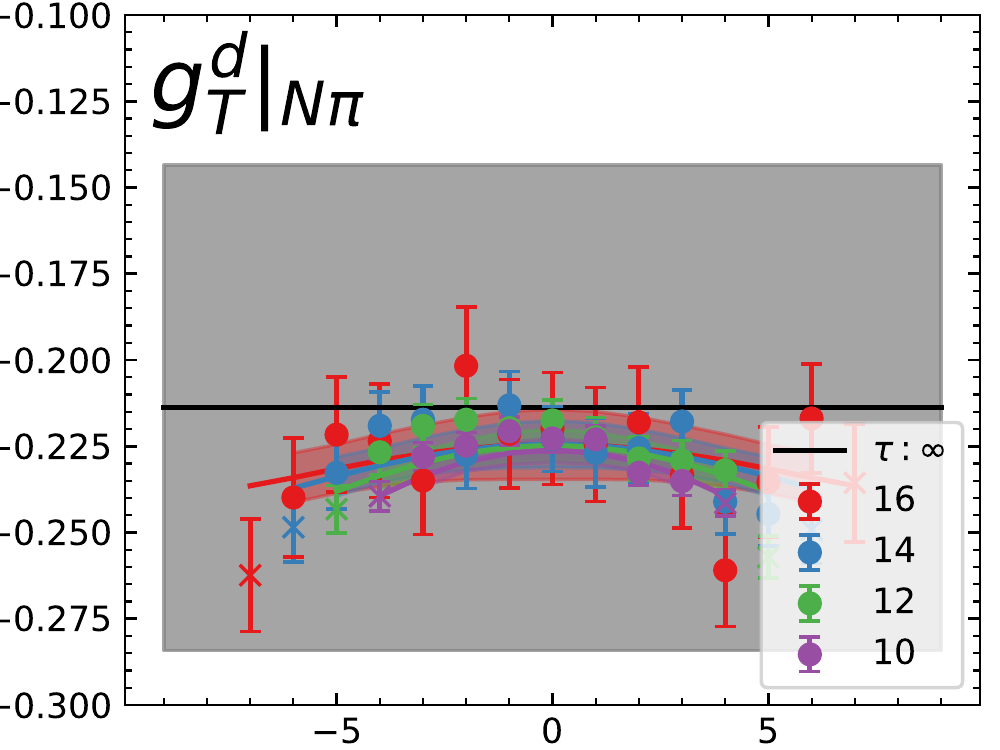}
      \phantom{\includegraphics[width=0.30\linewidth]{figs/3pt_AIC/gTd_4Npi_a09m130.pdf}}
      
      \includegraphics[width=0.30\linewidth]{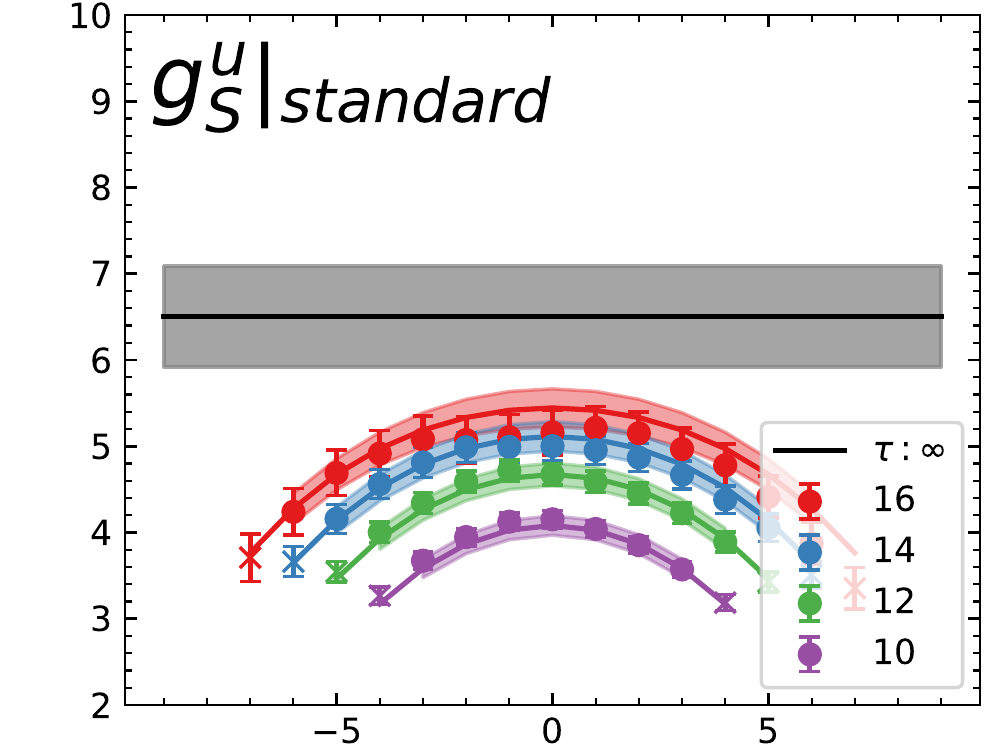}
      \includegraphics[width=0.30\linewidth]{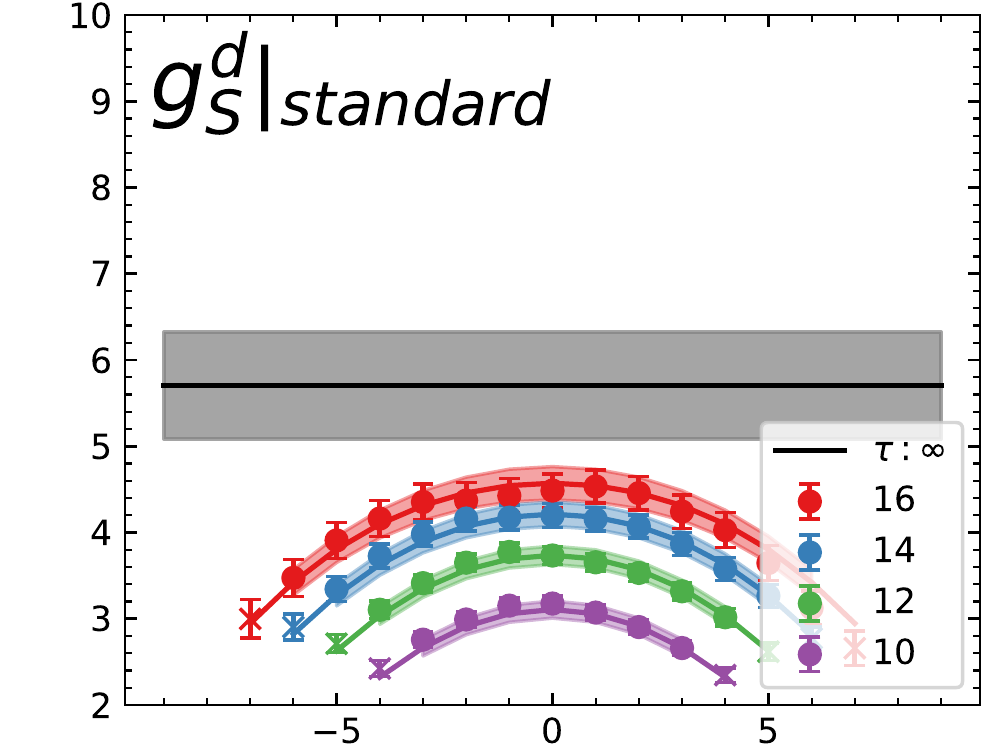}
      \includegraphics[width=0.30\linewidth]{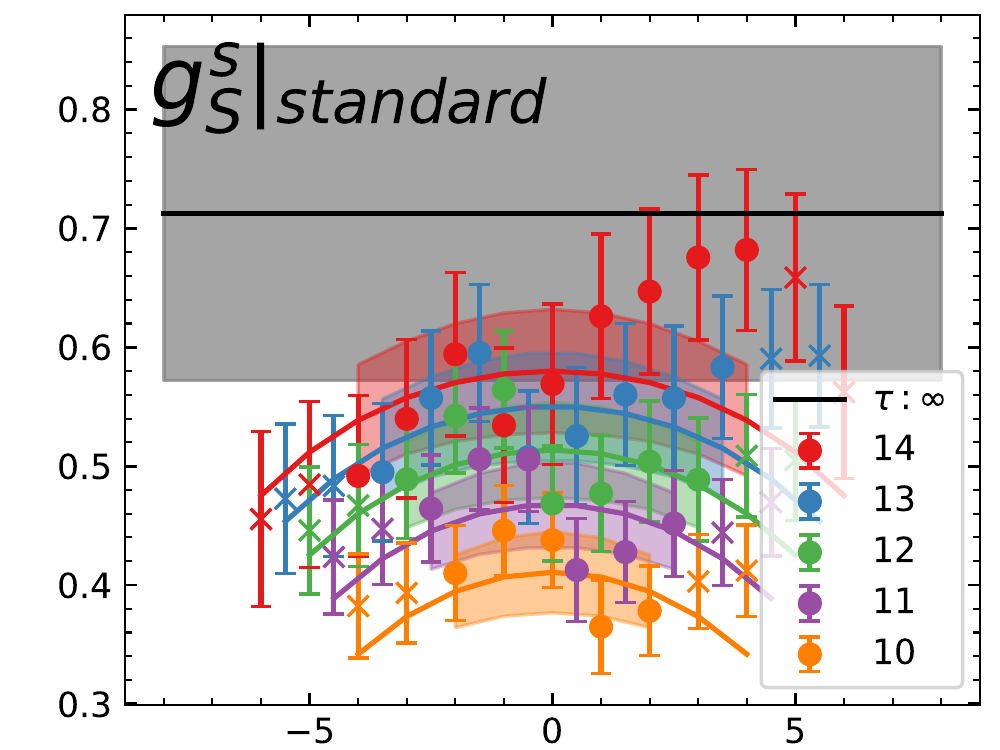}
            
      \includegraphics[width=0.30\linewidth]{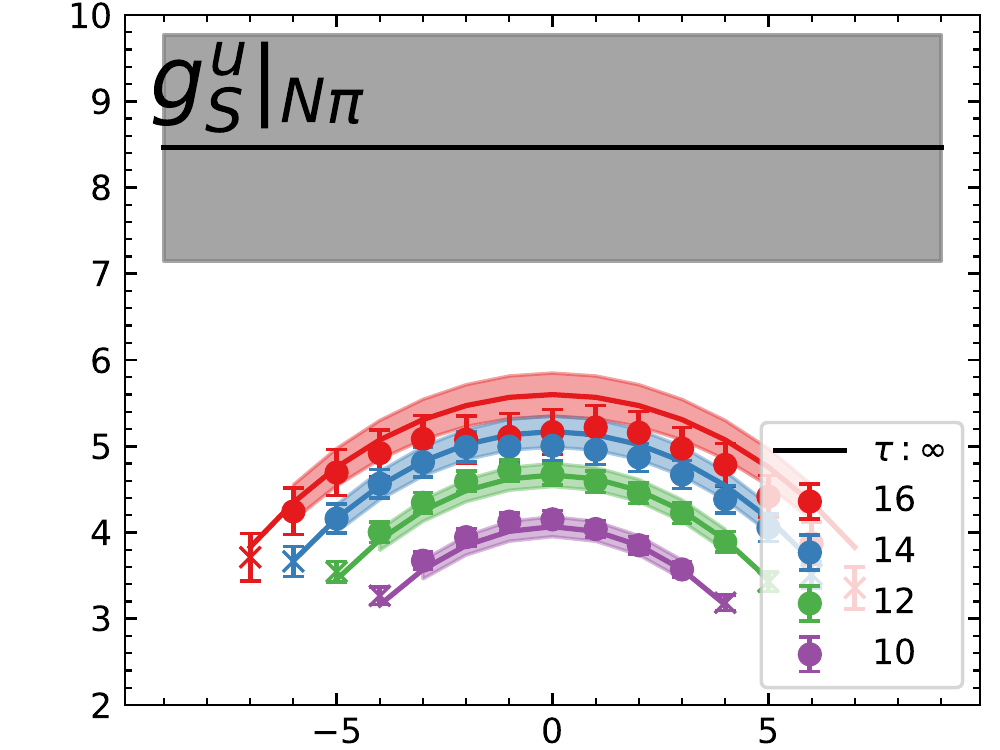}
      \includegraphics[width=0.30\linewidth]{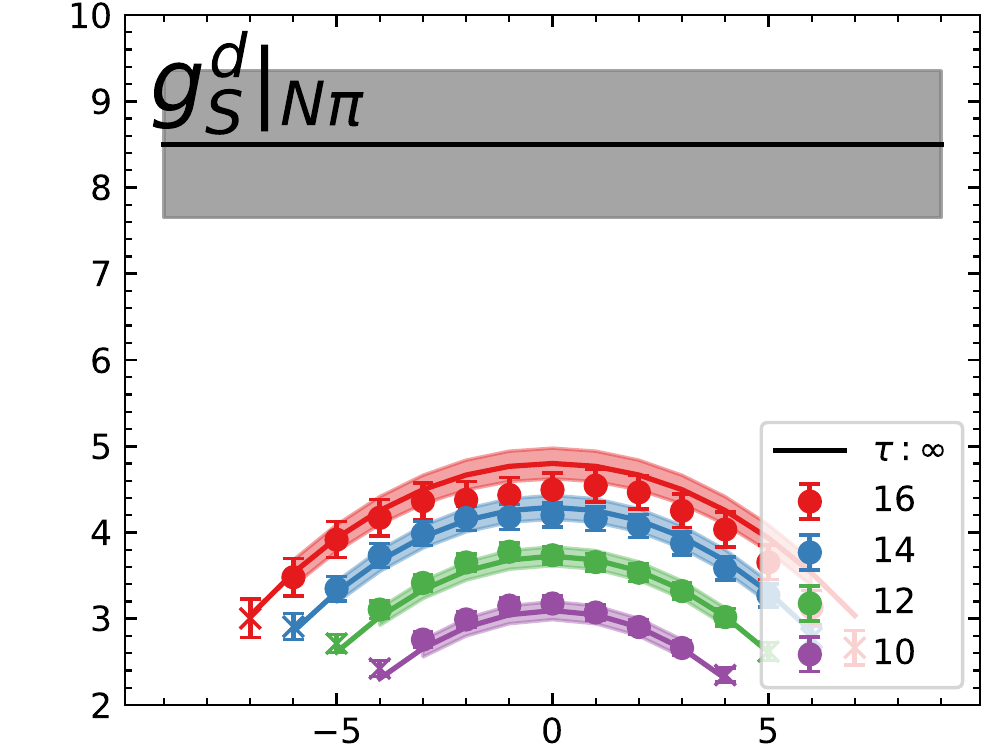}
      \phantom{\includegraphics[width=0.30\linewidth]{figs/3pt_AIC/gSd_4Npi_a09m130.pdf}}
  \caption{The physical $\mpi$ ensemble ($a09m130$) data for the sum of the 
  connected and disconnected contributions at 
  different $\{\tau,t\}$ 
  are plotted versus $(t-\tau/2)/a$.   Different panels show excited-state fits to data 
   for bare $g_A$, $g_T$, and $g_S$ using the ``standard'' and $N \pi$ strategies 
   defined in the text. Result of the fit is shown
    by lines of the same color as the data for various $\tau/a$ listed
    in the label, and the $\tau\to\infty$ value is given by the gray band. }
  \label{fig:gAgTgS_ESC}
\end{figure}

\section{Update on results for $g_{A,S,T}$}
\label{sec:results}

\begin{figure}[] 
  \centering
      \includegraphics[width=0.235\linewidth]{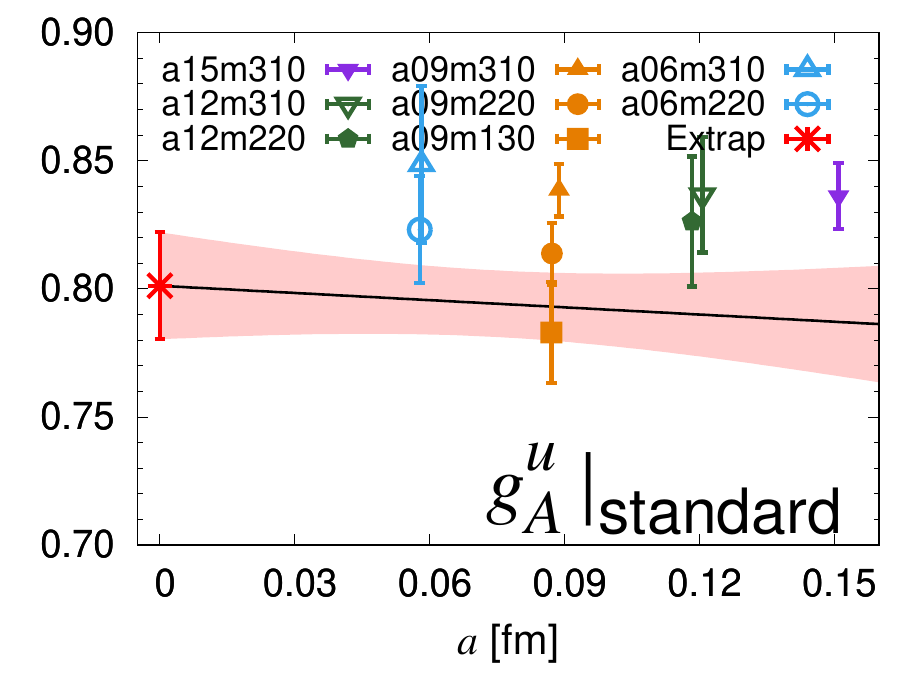}
      \includegraphics[width=0.235\linewidth]{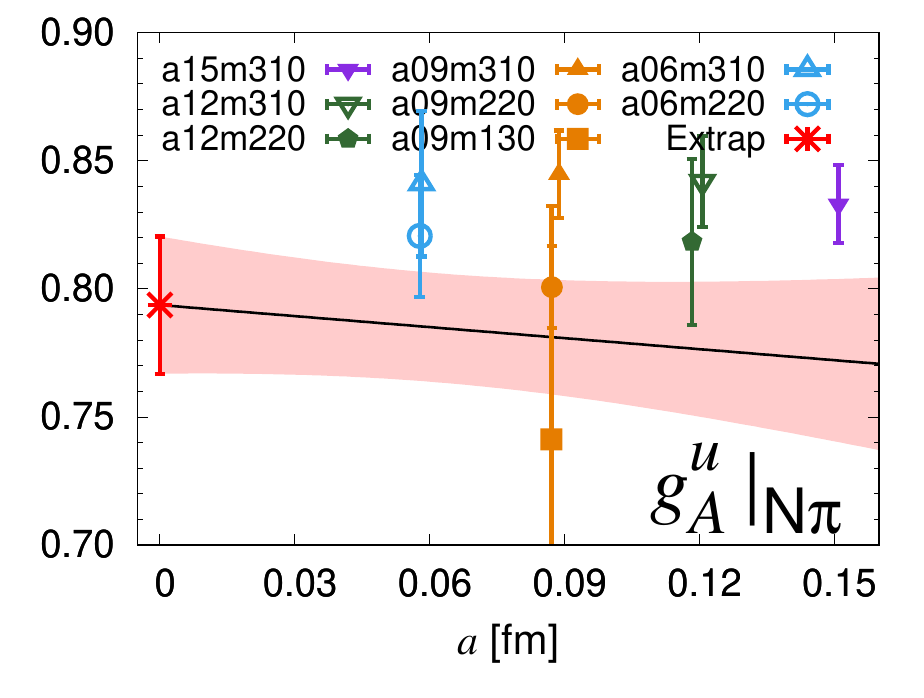}
      \includegraphics[width=0.235\linewidth]{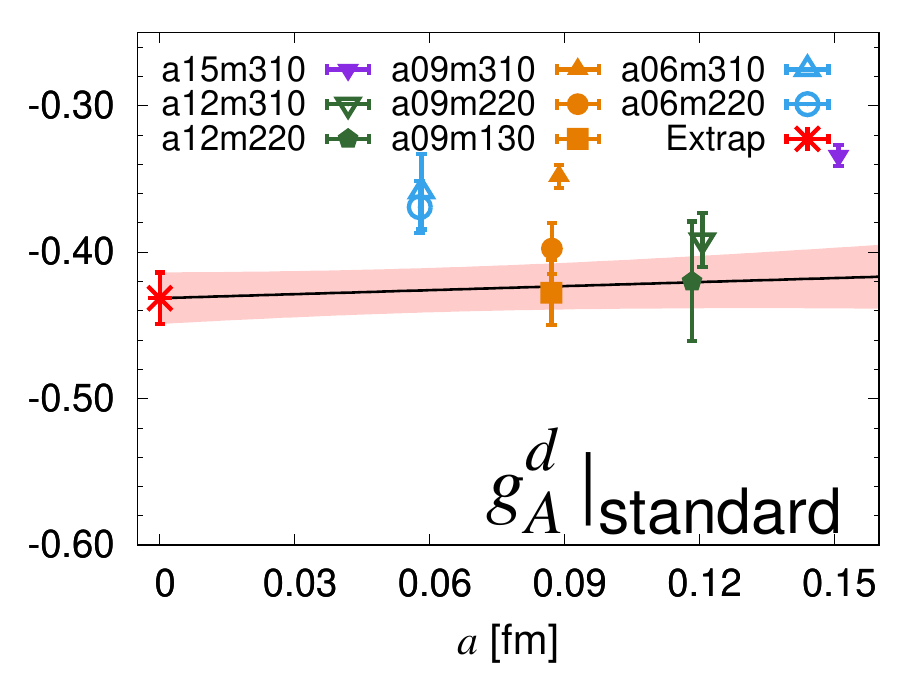}
      \includegraphics[width=0.235\linewidth]{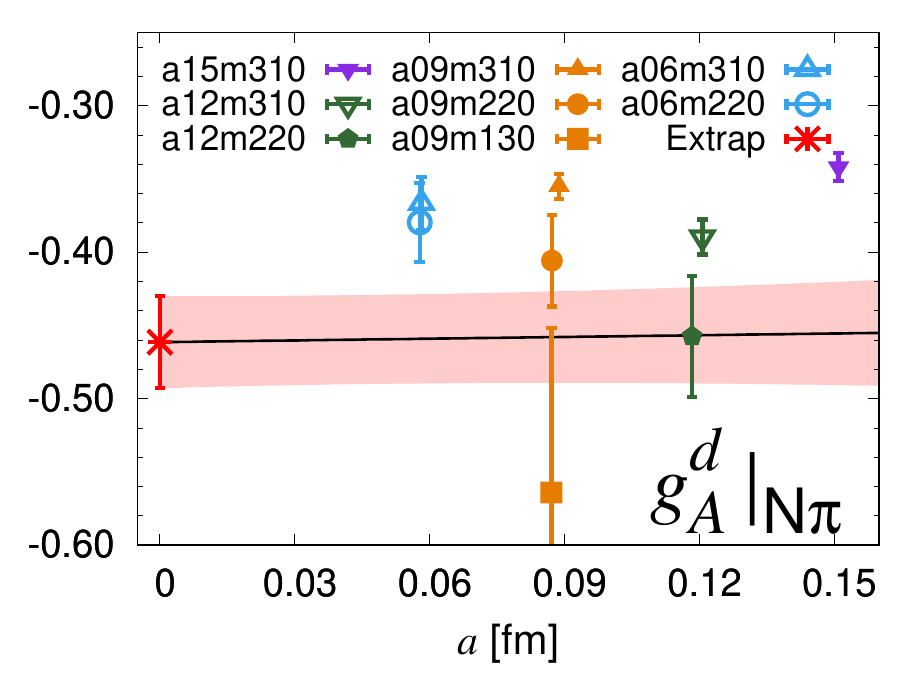}
  
      \includegraphics[width=0.235\linewidth]{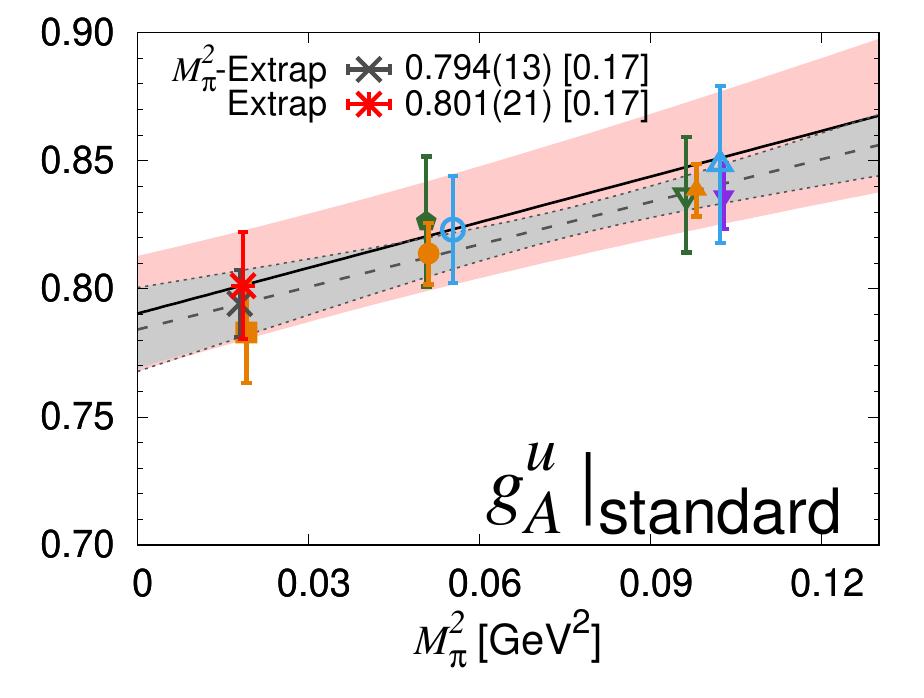}
      \includegraphics[width=0.235\linewidth]{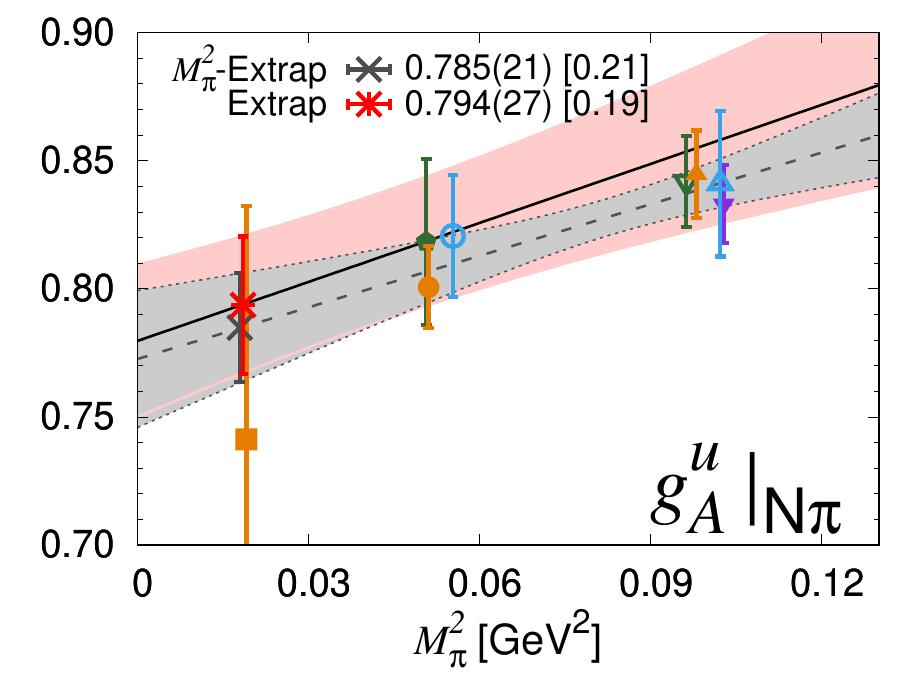}
      \includegraphics[width=0.235\linewidth]{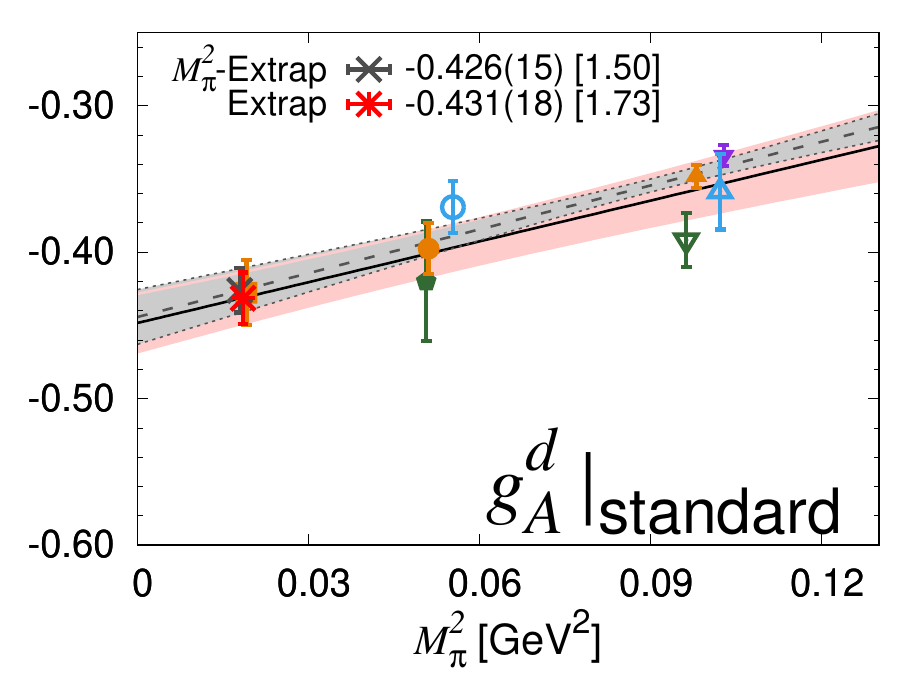}
      \includegraphics[width=0.235\linewidth]{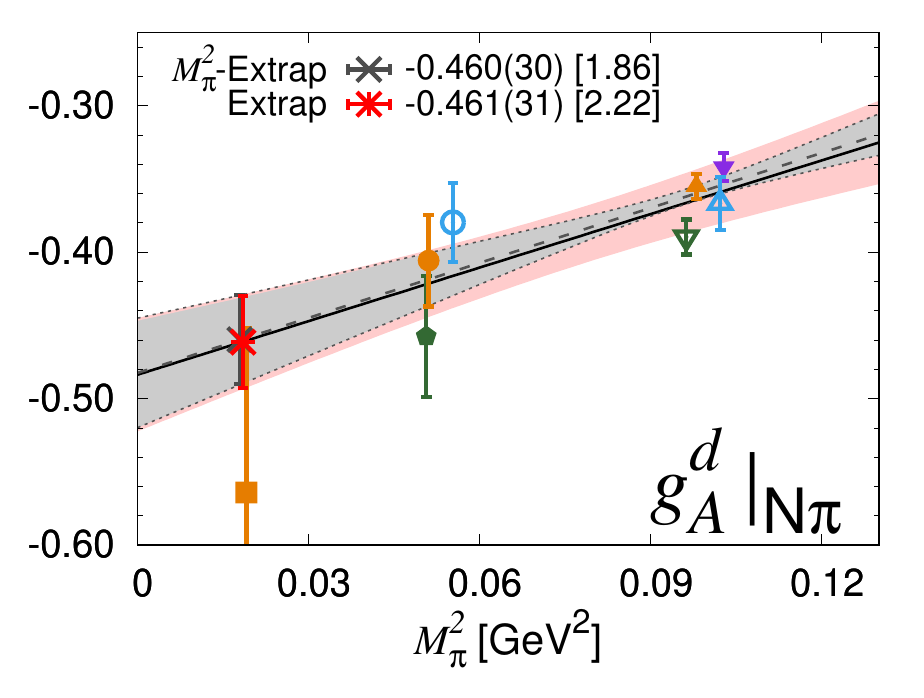}
  
  \vspace{-0.1in}
  \caption{CC fits to $g_{A}^{u}$ (left two panels) and $g_{A}^{d}$ (right two panels) obtained with standard and $N\pi$ strategies using the ansatz $d_0+d_a a+ d_2 M_\pi^2$. Fit result is plotted versus $a$ in top row and versus $M_\pi^2$ in bottom row.}
  \label{fig:CC_gA}
\end{figure}

\begin{figure}[] 
  \centering
      \includegraphics[width=0.235\linewidth]{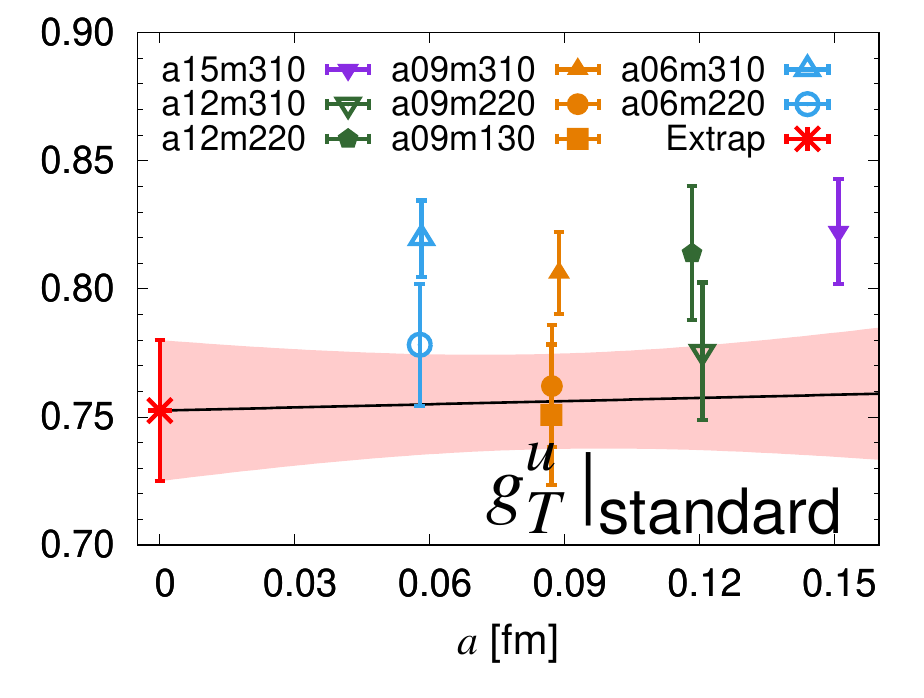}
      \includegraphics[width=0.235\linewidth]{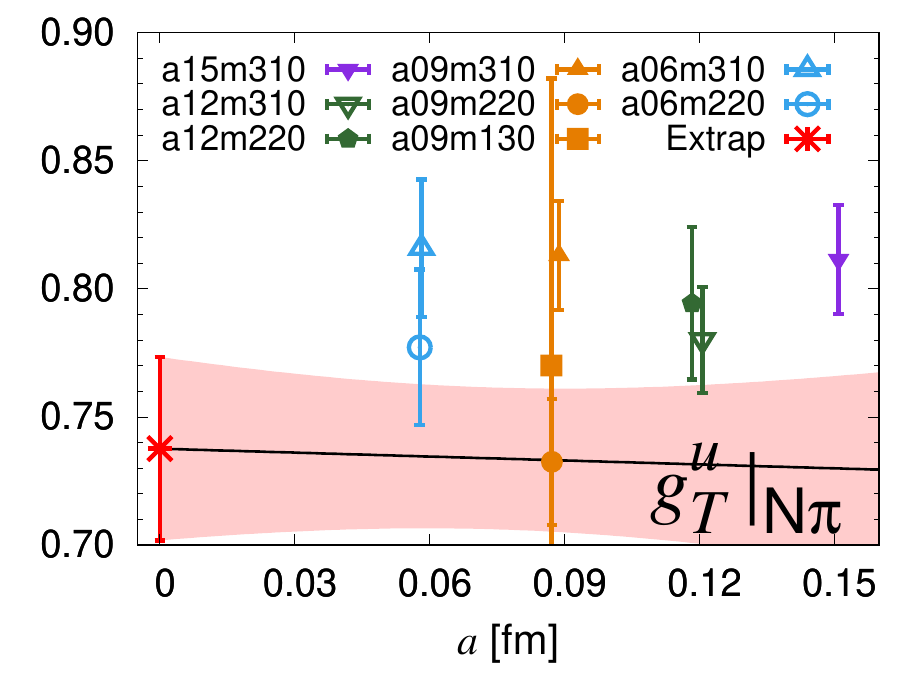}
      \includegraphics[width=0.235\linewidth]{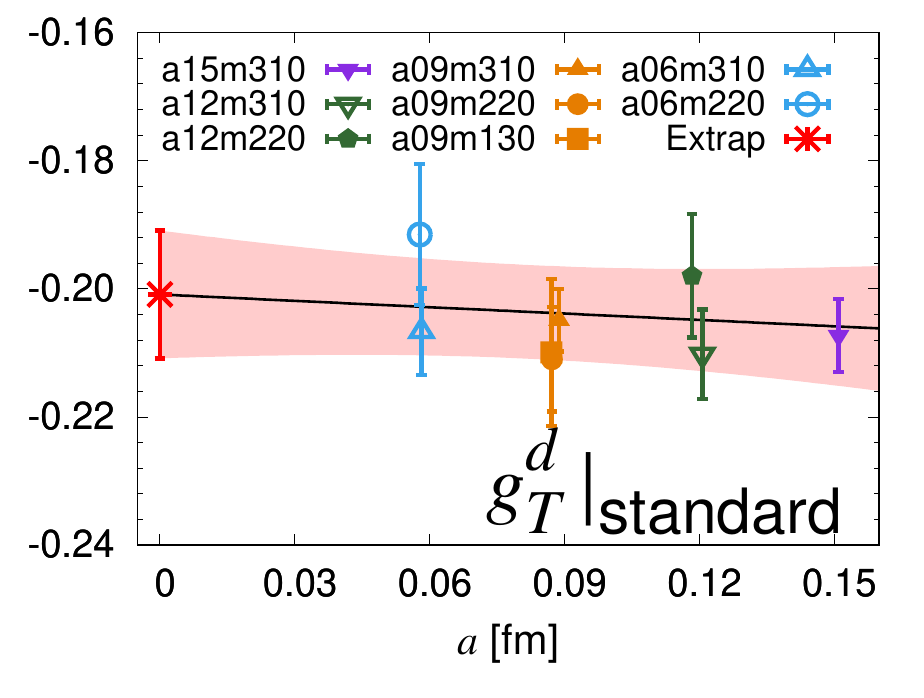}
      \includegraphics[width=0.235\linewidth]{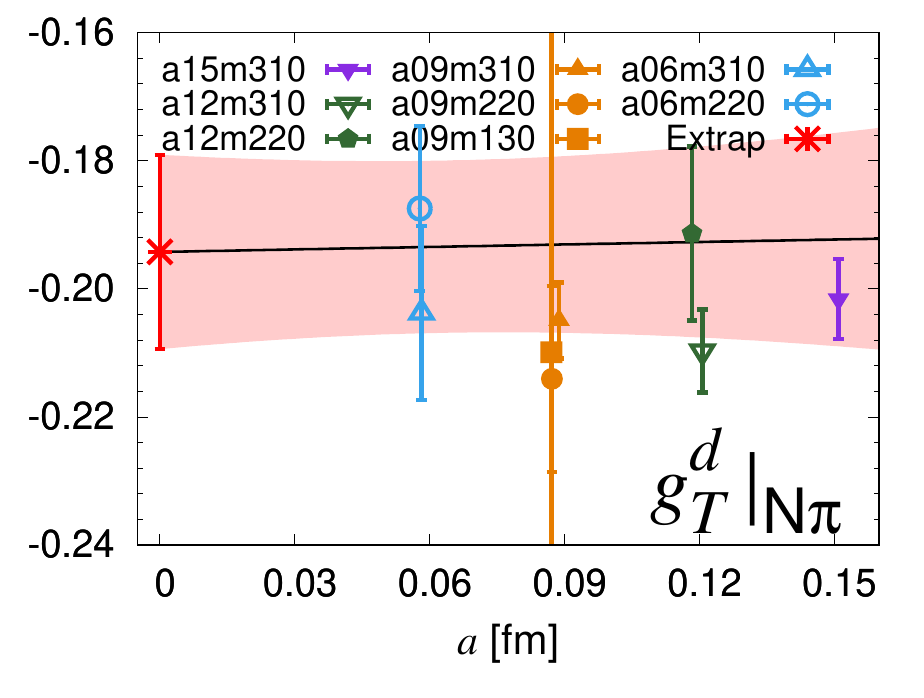}
  
      \includegraphics[width=0.235\linewidth]{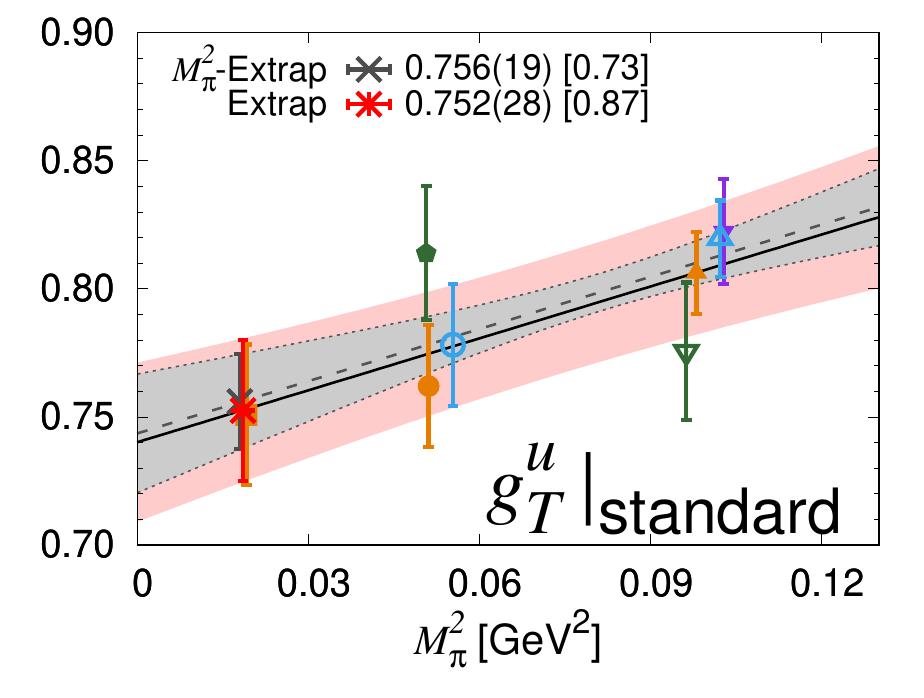}
      \includegraphics[width=0.235\linewidth]{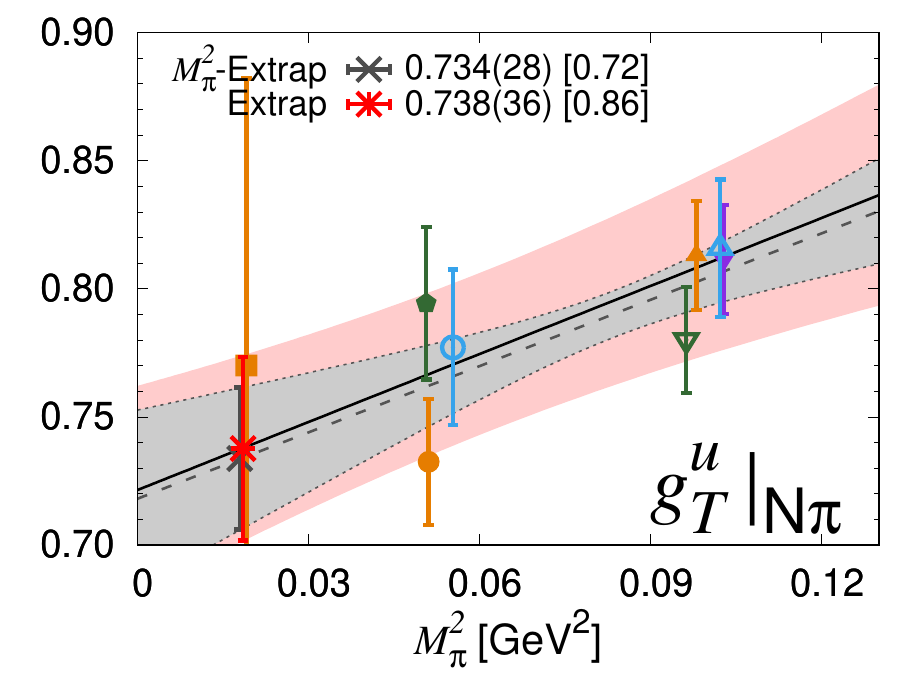}
      \includegraphics[width=0.235\linewidth]{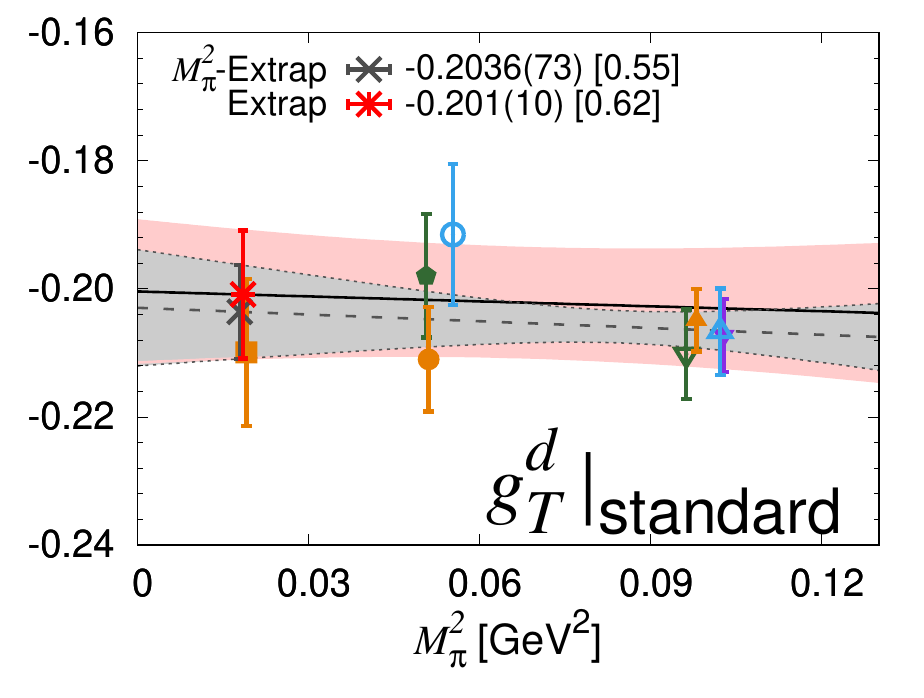}
      \includegraphics[width=0.235\linewidth]{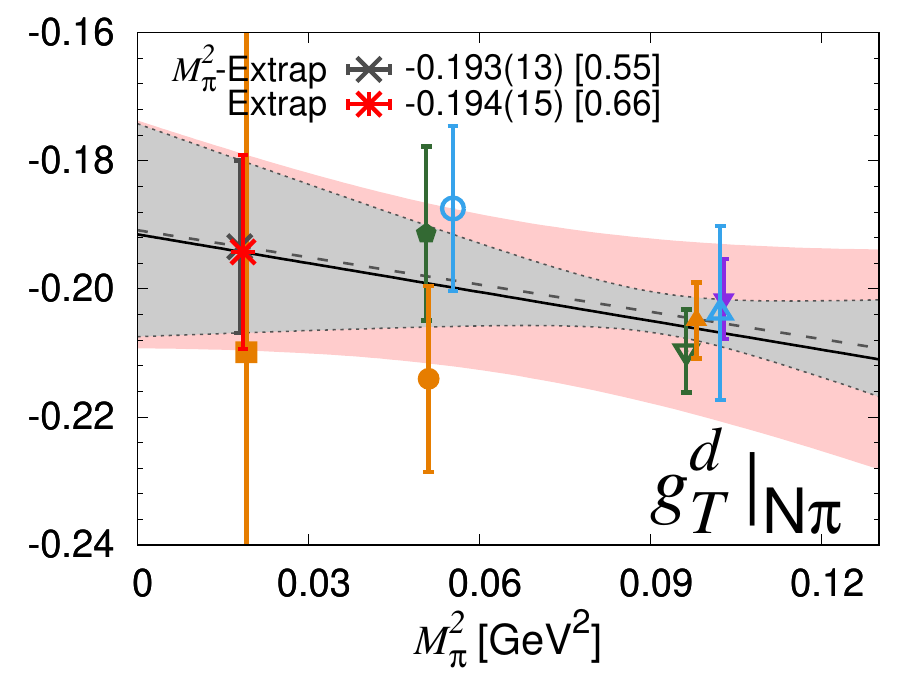}
  
  \vspace{-0.1in}
  \caption{CC fits to $g_{T}^{u}$ (left two panels) and $g_{T}^{d}$ (right two panels) obtained with standard and $N\pi$ strategies using the ansatz $d_0+d_a a+ d_2 M_\pi^2$. Fit result is plotted versus $a$ in top row and versus $M_\pi^2$ in bottom row.}
  \label{fig:CC_gT}
\end{figure}

\begin{figure}[] 
  \centering
      \includegraphics[width=0.235\linewidth]{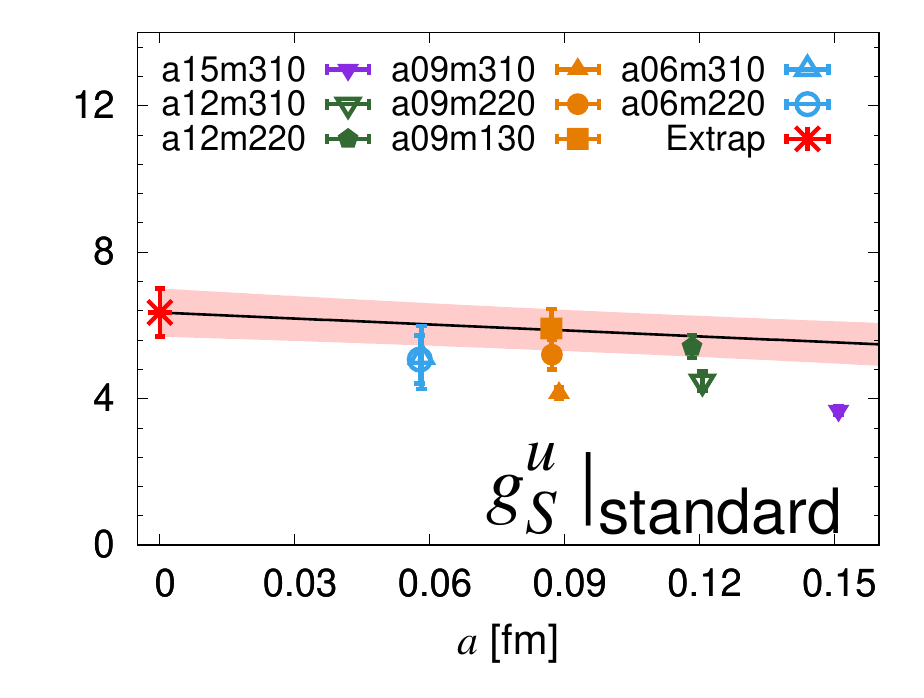}
      \includegraphics[width=0.235\linewidth]{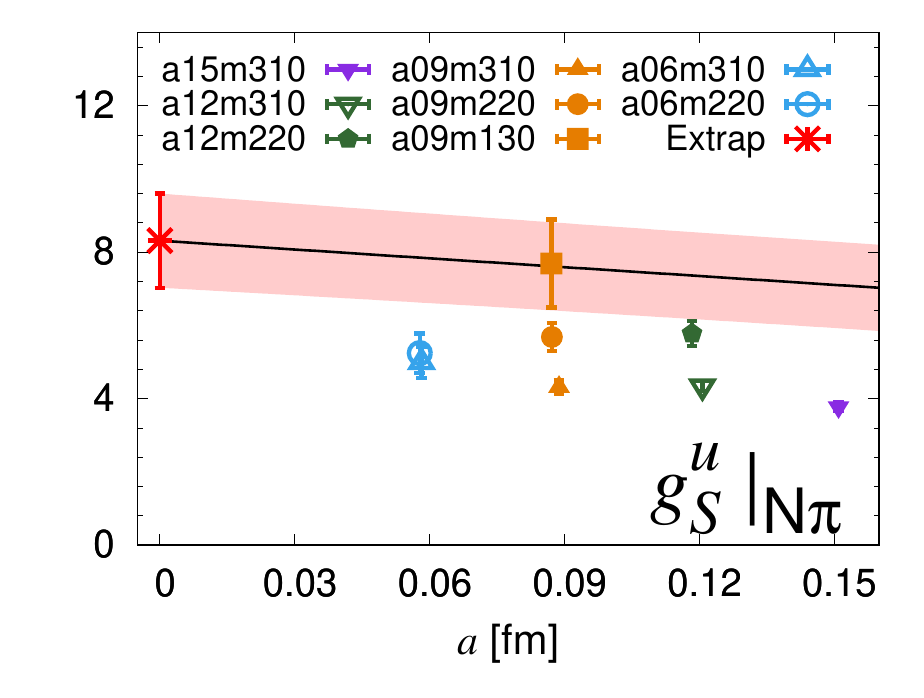}
      \includegraphics[width=0.235\linewidth]{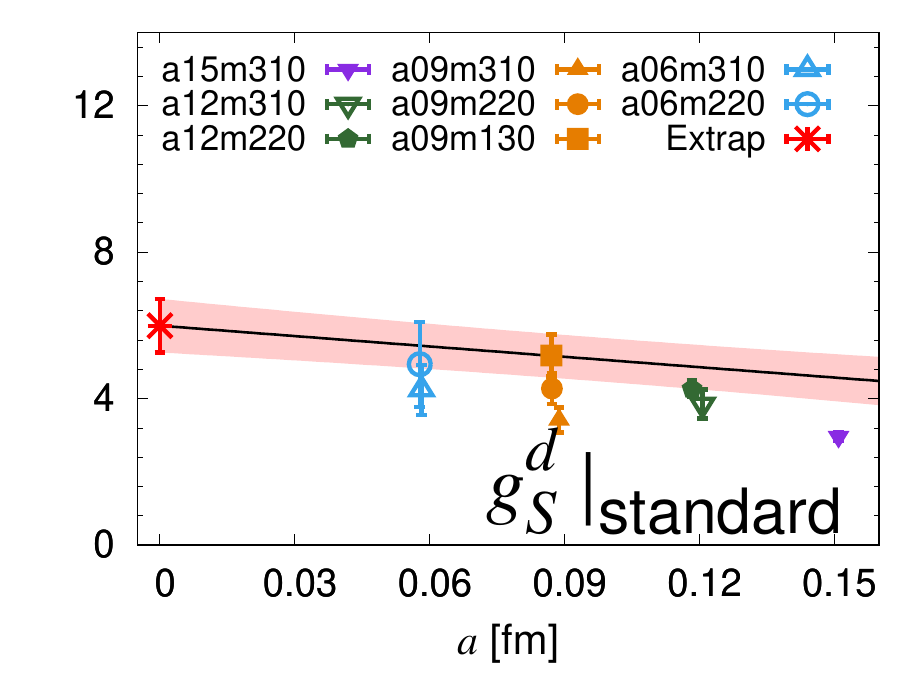}
      \includegraphics[width=0.235\linewidth]{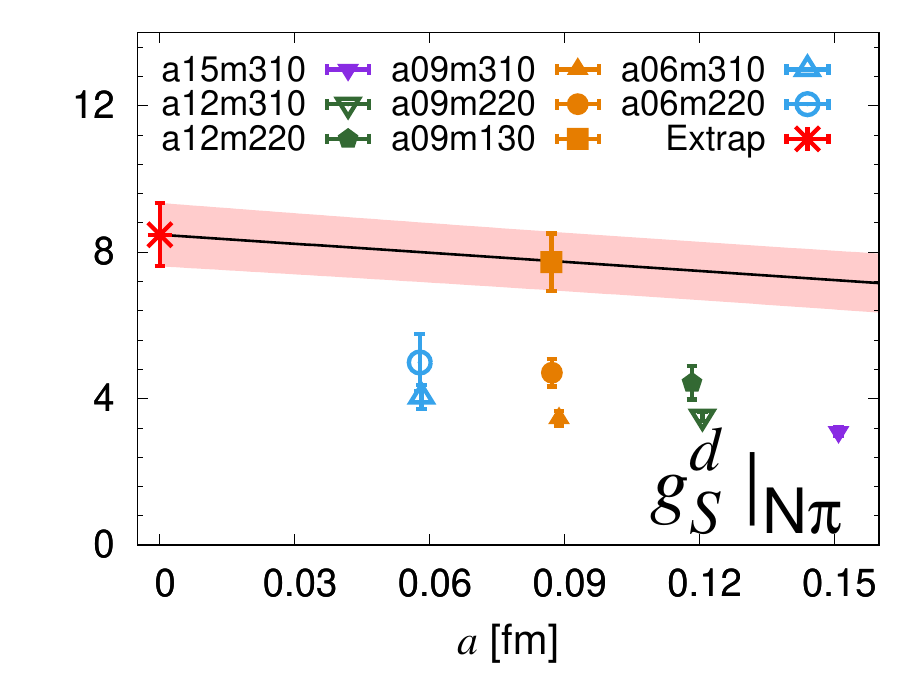}
  
      \includegraphics[width=0.235\linewidth]{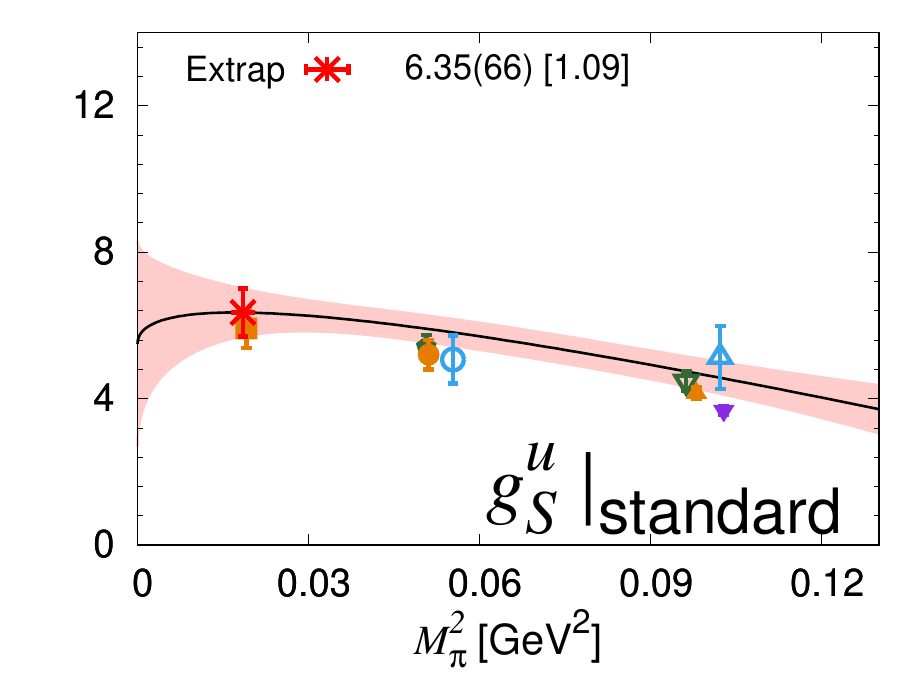}
      \includegraphics[width=0.235\linewidth]{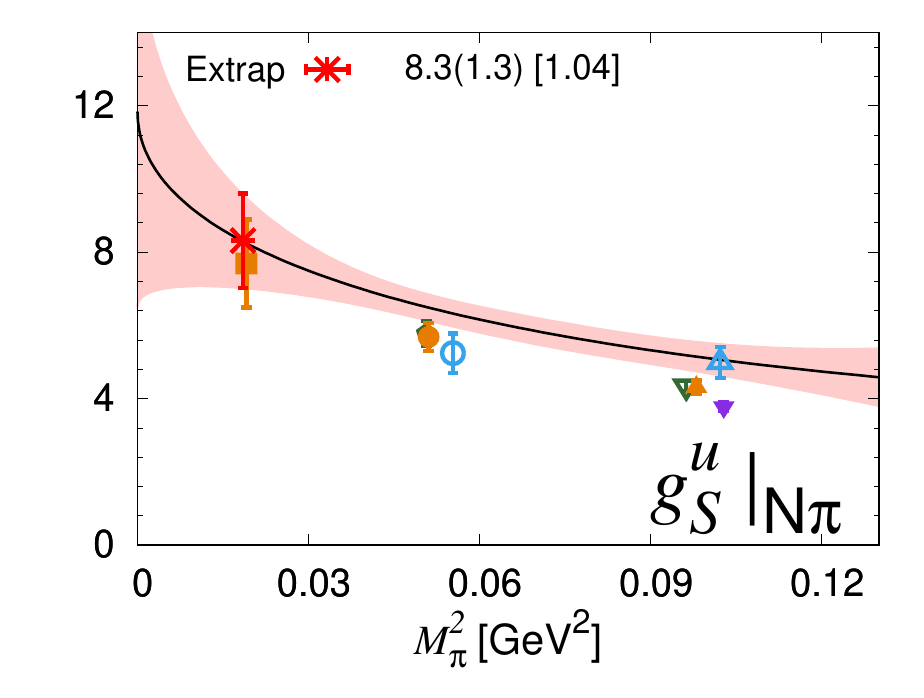}
      \includegraphics[width=0.235\linewidth]{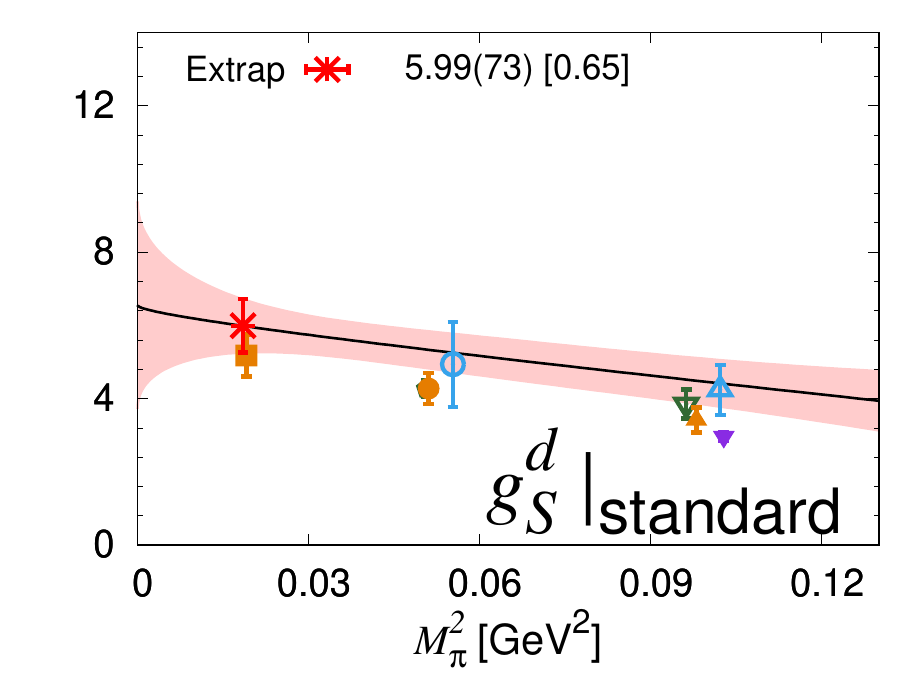}
      \includegraphics[width=0.235\linewidth]{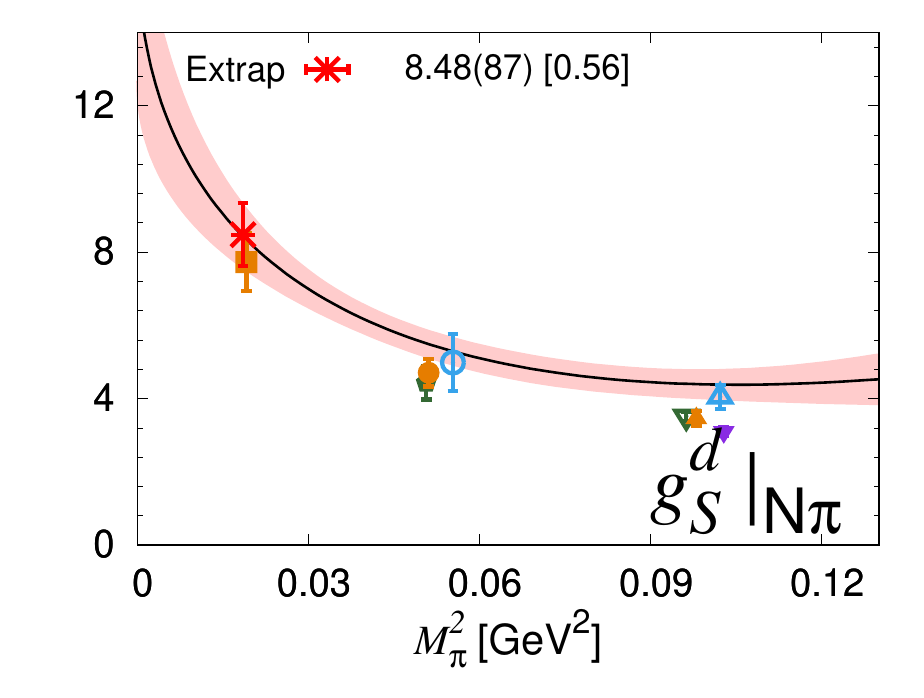}
  
  \vspace{-0.1in}
  \caption{CC fits to $g_{S}^{u}$ (left 2 panels) and $g_{S}^{d}$ (right 2 panels) obtained with standard and $N\pi$ strategies using the ansatz $d_0+d_a a+ d_1 M_\pi+ d_2 M_\pi^2$ to $g_{S}^{u,d}$. Fit result is plotted versus $a$ in top row and versus $M_\pi^2$ in bottom row.\looseness-1}
  \label{fig:CC_gS}
\end{figure}

\begin{figure}[] 
  \centering
  \includegraphics[width=0.32\linewidth]{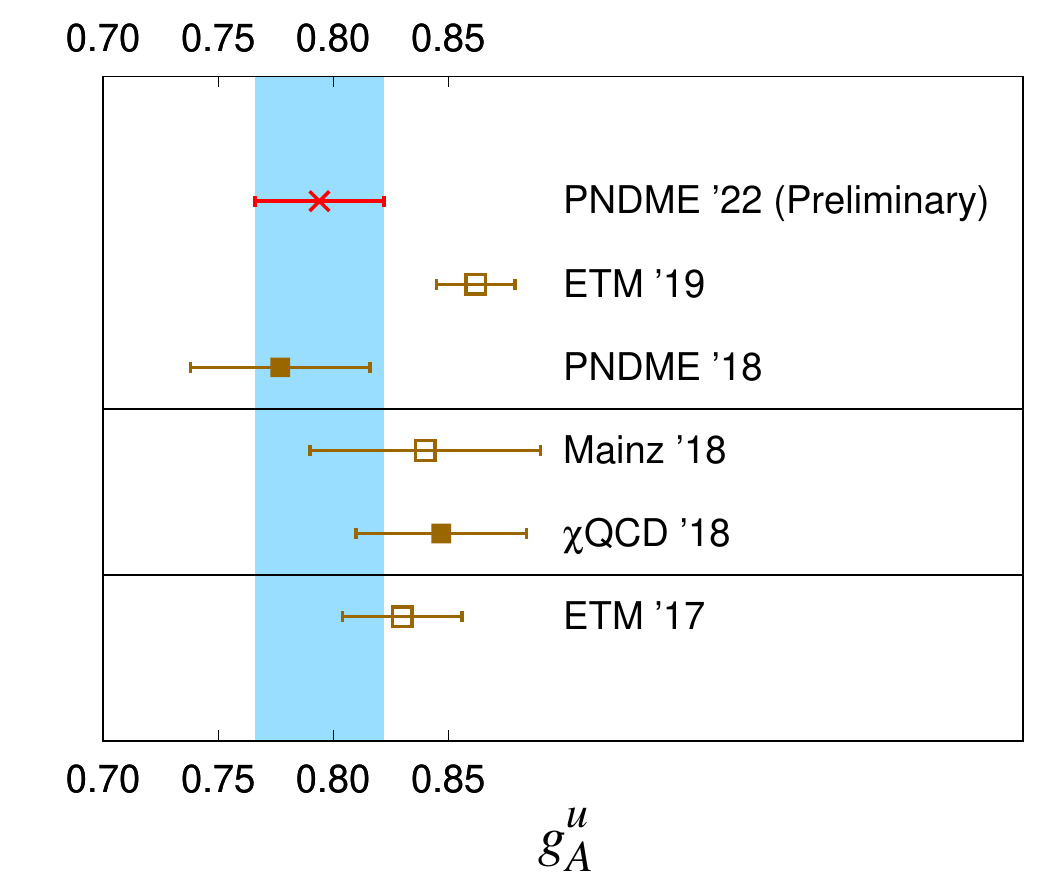}
  \includegraphics[width=0.32\linewidth]{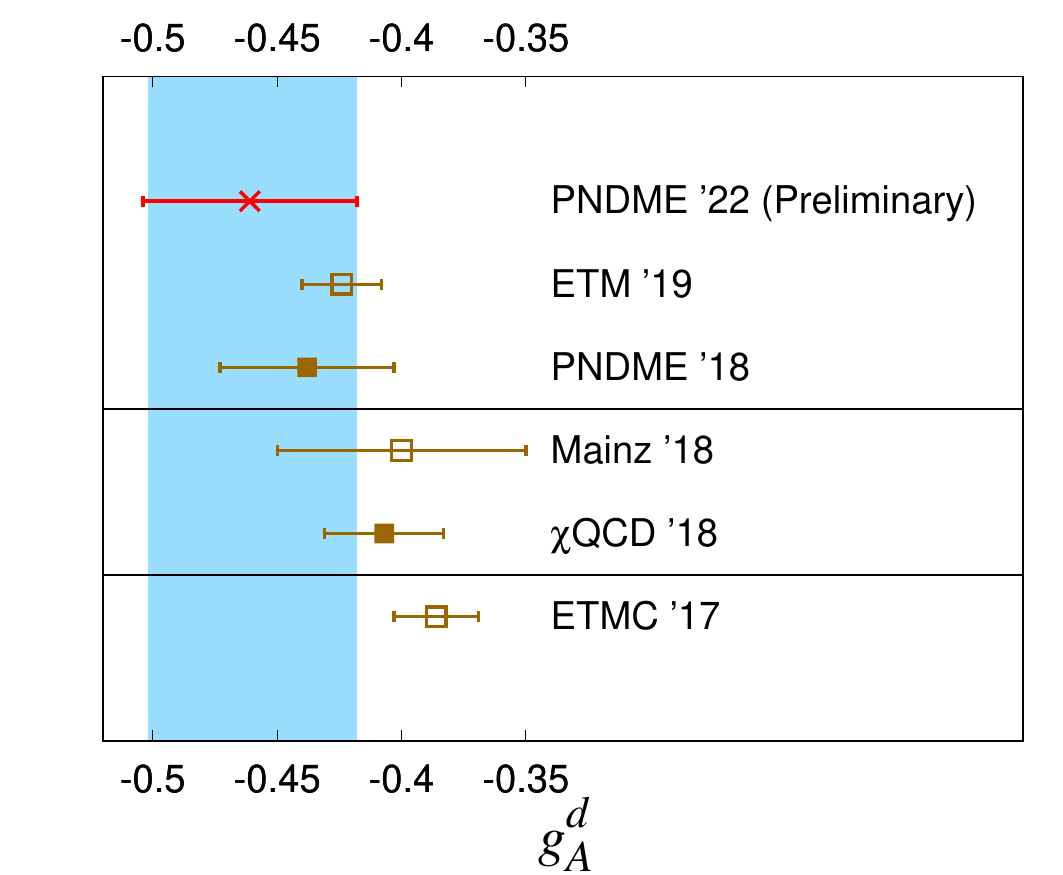}
  \includegraphics[width=0.32\linewidth]{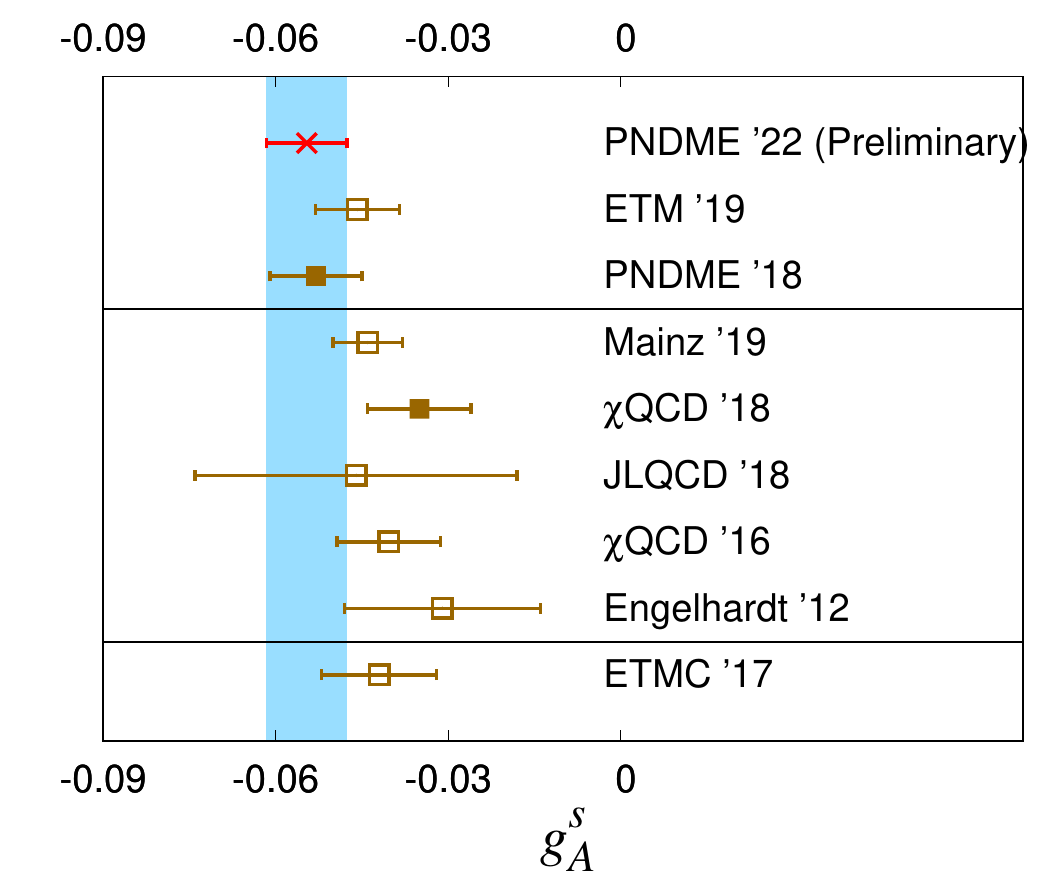}
  
  \includegraphics[width=0.32\linewidth]{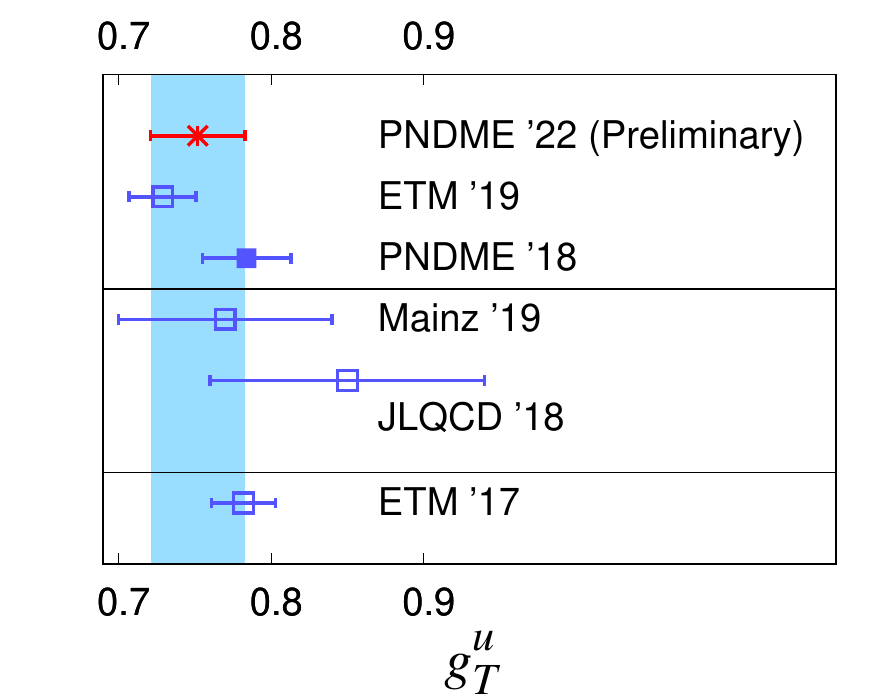}
  \includegraphics[width=0.32\linewidth]{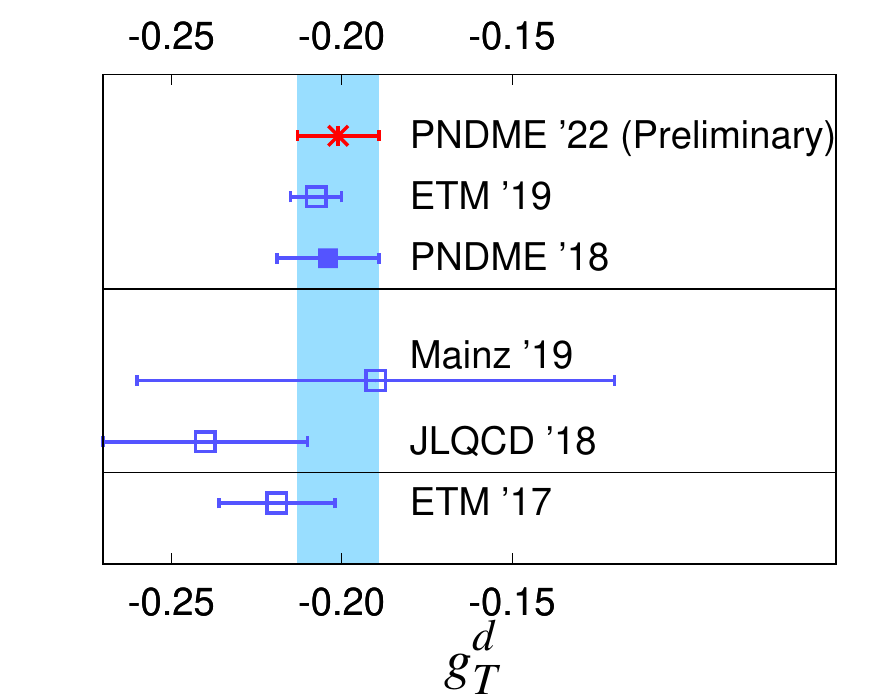}
  \includegraphics[width=0.32\linewidth]{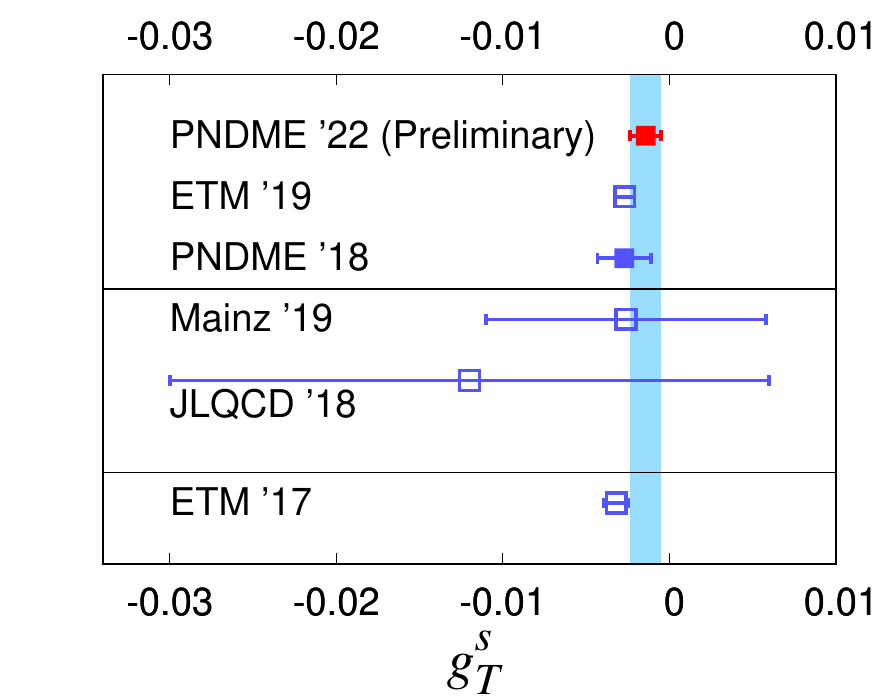}
  \caption{Preliminary (PNDME 22) and published PNDME 18~\cite{Gupta:2018lvp,Lin:2018obj} results for $g_A^q$ (top) and $g_T^q$ (bottom) added to the FLAG 2021 summary figure. See Ref.~\cite{FLAG:2021npn} for details and references to other works. }
  \label{fig:flag_gAgT}
\end{figure}

\begin{figure}[]  
  \centering
  \begin{subfigure}[$\sigma_{\pi N}$]{\includegraphics[width=0.48\linewidth]{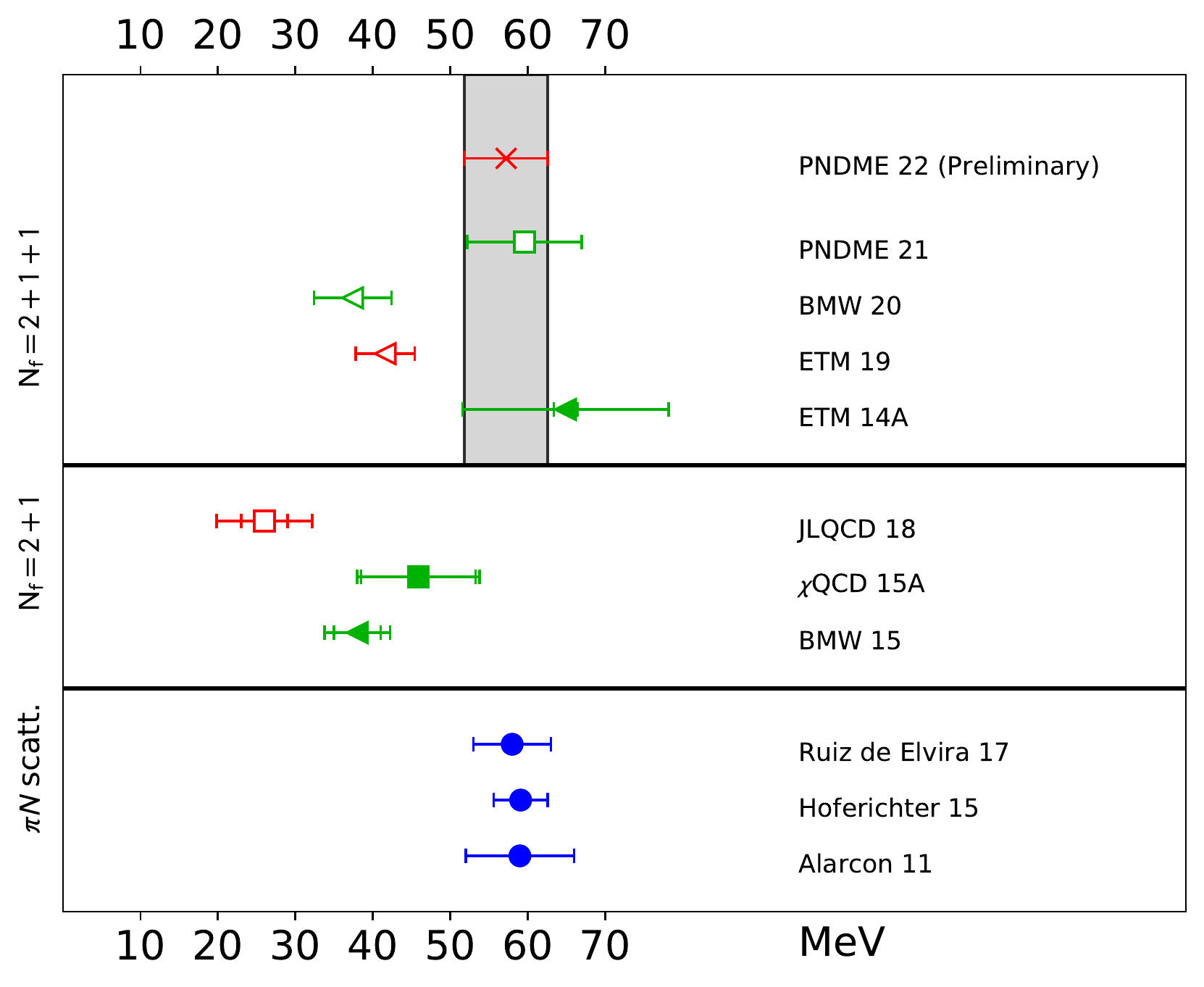}}\end{subfigure}
  \begin{subfigure}[$\sigma_{s}$]{\includegraphics[width=0.48\linewidth]{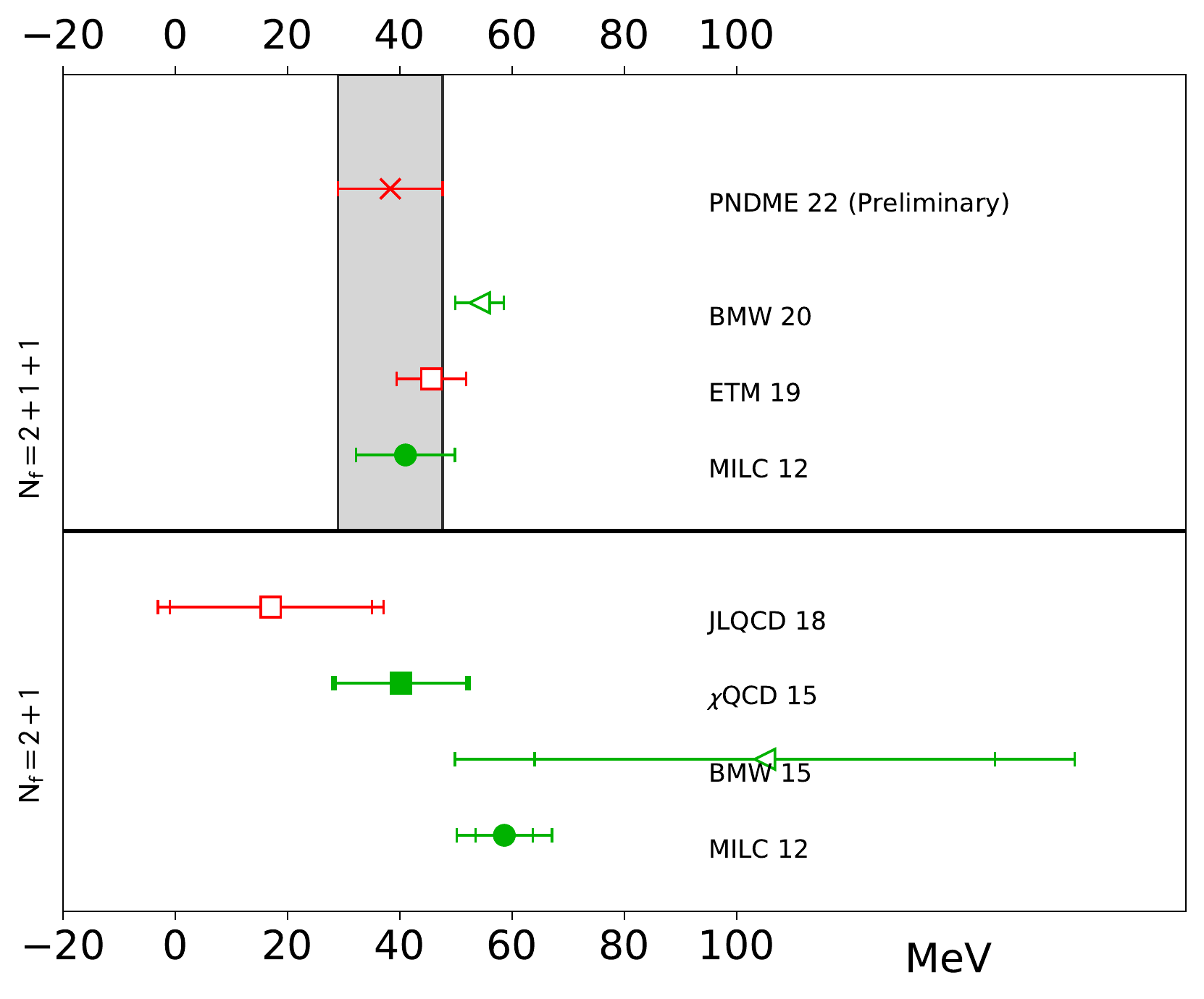}}\end{subfigure}
  \caption{Our results (PNDME 22) and published PNDME 21~\cite{Gupta:2021ahb}) for the pion-nucleon sigma term and the strangeness content of the nucleon. For details and references to other lattice calculations, see Ref.~\cite{Gupta:2021ahb,FLAG:2021npn}.  }
  \label{fig:flag_sigma}
\end{figure}

Examples of ESC fits to $a091m130$ data are shown in Fig.~\ref{fig:gAgTgS_ESC}, with the largest difference between the two strategies observed in $g_{S}$.  For the charges with strange flavor,
$g_{A,T,S}^s$, the leading multihadron excited state is expected to be
$\Sigma K$, which has a large mass gap, so we consider the
``standard'' analysis more appropriate for it. The chiral-continuum
(CC) fits to the renormalized charges are shown in
Figs.~\ref{fig:CC_gA},~\ref{fig:CC_gT} and~\ref{fig:CC_gS}.  Possible
finite-volume corrections are ignored in this analysis.  The final preliminary results
are summarized in Fig.~\ref{fig:flag_gAgT} and~\ref{fig:flag_sigma} and Table~\ref{tab:gAST}.
Some details are as follows:

{\textbf{Axial charges, $g_A^{u,d,s}$}}:

Compared to the ``standard'' analysis for $g_A^{u,d}$, 
the ``$N\pi$'' analysis finds larger ESC especially in the physical smaller pion mass ensemble as shown in Fig.~\ref{fig:gAgTgS_ESC}.
CC fits in Fig.~\ref{fig:CC_gA} for  
both $g_A^u$ and $g_A^d$ show similar dependence on $a$ and
$\mpi$, with larger uncertainty in the $N\pi$ analysis. There is a significant slope versus $M_\pi^2$ that adds for $g_A^{u+d}$ and almost cancels for $g_A^{u-d}$. $g_A^u$ is almost
unchanged with or without ``$N\pi$'', while $|g_A^d|$ is $\approx 7\%$
larger with ``$N\pi$'' analysis. Our final extrapolated $g_A^q$ with the ``$N\pi$'' analysis of ESC (as motivated by the isovector axial form factor analysis described in \cite{Park:2021ypf}) are
summarized in Fig.~\ref{fig:flag_gAgT} (and Table~\ref{tab:gAST}) along with 
determinations from other collaborations taken from the FLAG review 2021 \cite{FLAG:2021npn}. The difference  between the extrapolated values with the ``standard'' and
``$N\pi$'' analysis data can be viewed as an additional systematic error. Results for $g_A^{u,d,s}$ are consistent with those published in Ref.~\cite{Lin:2018obj}.

{\textbf{Tensor charges, $g_T^{u,d,s}$}}:
The magnitude of ESC in $g_T^u$ and $g_T^d$ is similar. The ``standard'' and ``$N\pi$'' analysis of ESC, shown in Fig.~\ref{fig:gAgTgS_ESC}, give consistent central values. 
The statistical quality of the data on the physical pion mass ensemble, $a09m130$, is poor as shown in Fig.~\ref{fig:gAgTgS_ESC} and Ref.~\cite{Gupta:2018lvp}. In $g_T^s$, when 
there is no clear ESC pattern in $C^{\text{3pt},s}_T(t;\tau)$, 
the central value is taken from a constant fit to the middle points. 
The CC fits are shown in Fig.~\ref{fig:CC_gT}. The extrapolated $g_T^q$, using the ``standard''
analysis of ESC, are summarized in the
Fig.~\ref{fig:flag_gAgT} and Table~\ref{tab:gAST}. They are consistent with those in
Ref.~\cite{Gupta:2018lvp}. \looseness-1

\begin{figure}[] 
  \centering
      \includegraphics[width=0.30\linewidth]{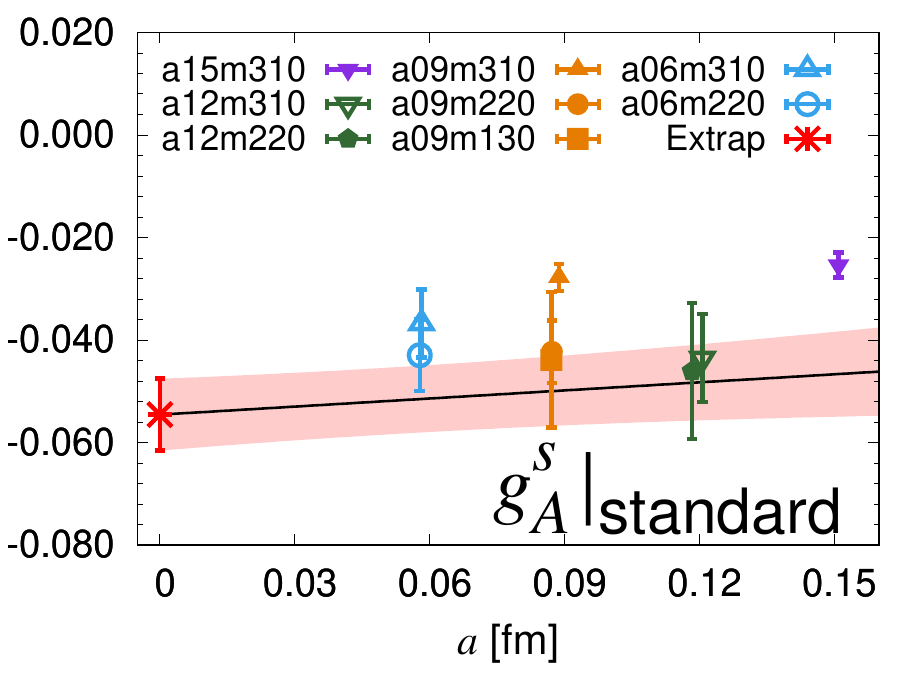}
      \includegraphics[width=0.30\linewidth]{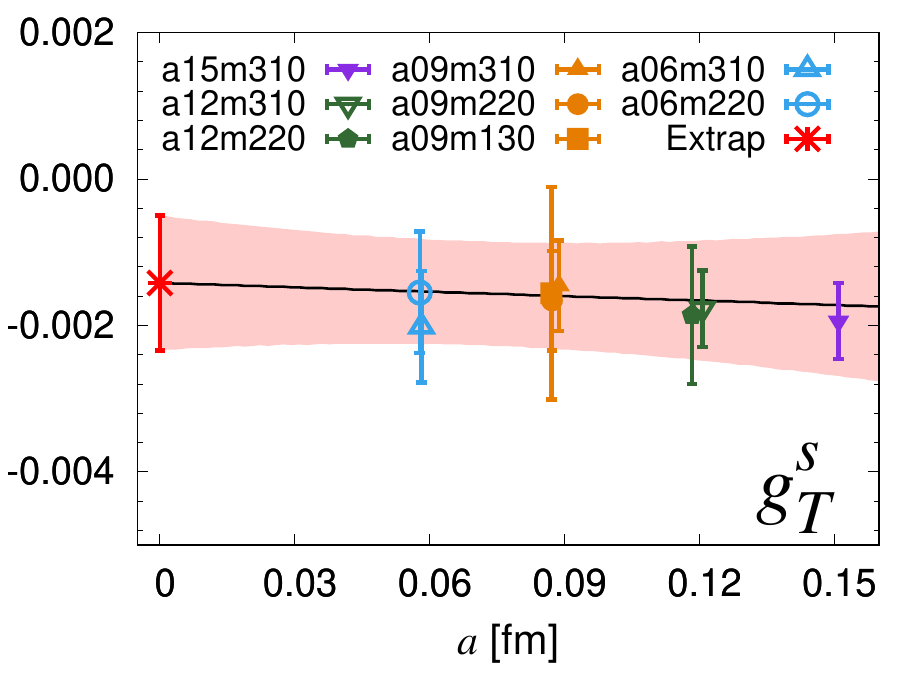}
      \includegraphics[width=0.30\linewidth]{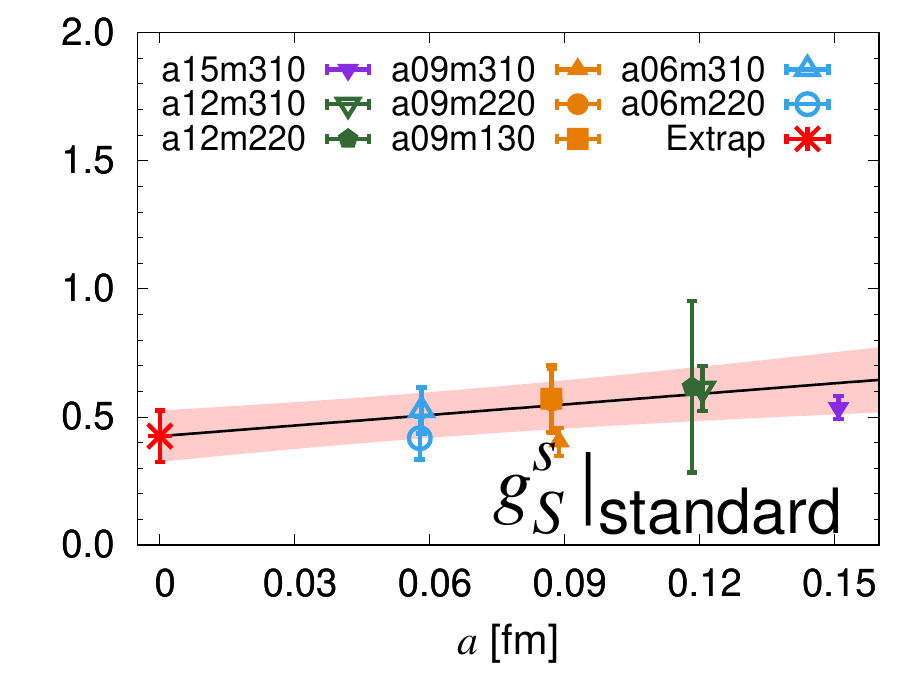}
      
      \includegraphics[width=0.30\linewidth]{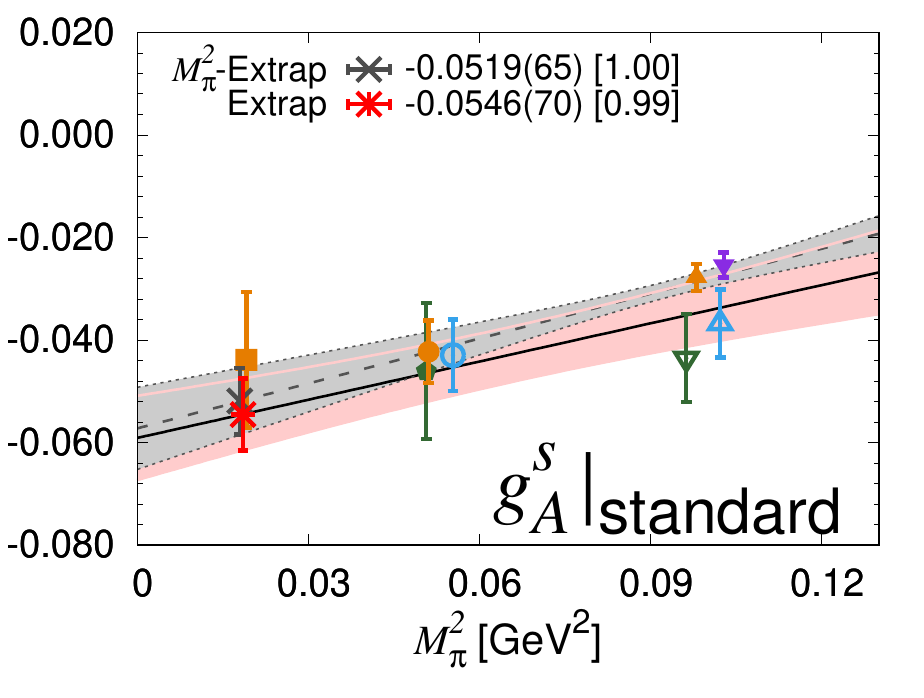}
      \includegraphics[width=0.30\linewidth]{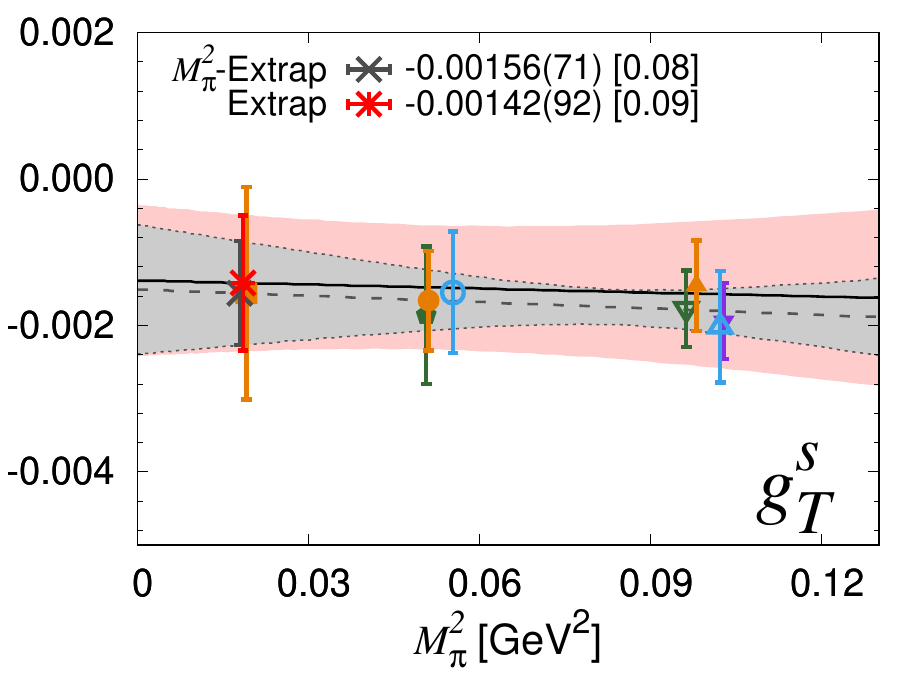}
      \includegraphics[width=0.30\linewidth]{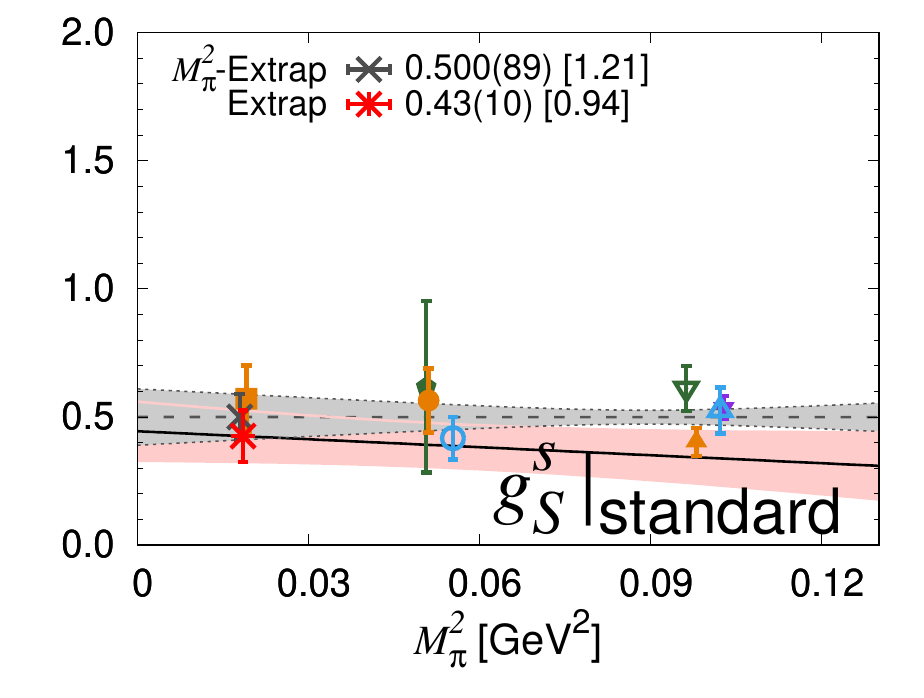}
  \vspace{-0.1in}
   \caption{CC fits to $g_{A,T,S}^{s}$ using the ansatz $d_0+d_a a+ d_2 M_\pi^2$ to $g_{S}^{s}$. We neglect dependence on $M_K$ since
   $m_s$ is tuned to its physical value.}

  \label{fig:CC_gASTs}
\end{figure}


\textbf{Scalar charges $g_S^{u,d,s}$:} 
Chiral PT gives two differences in the chiral behavior of
flavor diagonal scalar charges. First, the CC ansatz $g_S^{u,d} = 
d_0 + d_a a + d_1 \mpi + d_2 \mpi^2 + M_\pi^2 \log M_\pi^2 + \ldots$, has the chiral behavior starting
with a term proportional to $\mpi$~\cite{Hoferichter:2015hva}. Second, the
contribution of $N \pi$ and $N \pi \pi $ excited states is 
large in $g_S^{u,d}$~\cite{Gupta:2021ahb} as observed in the data shown in Fig.~\ref{fig:gAgTgS_ESC}.
The CC fits are shown in Fig.~\ref{fig:CC_gS}.   Our current estimates of $g_S^q$ are summarized
in Table~\ref{tab:gAST}. For $g_S^{u,d}$, the quoted results are from
the ``$N\pi$'' analysis motivated by Ref.~\cite{Gupta:2021ahb}, and
for $g_S^s$, we use the ``standard'' analyses as the lowest multihadron state should be the $\Sigma K$.

\begin{figure}[p] 
  \centering

  \includegraphics[width=0.24\linewidth]{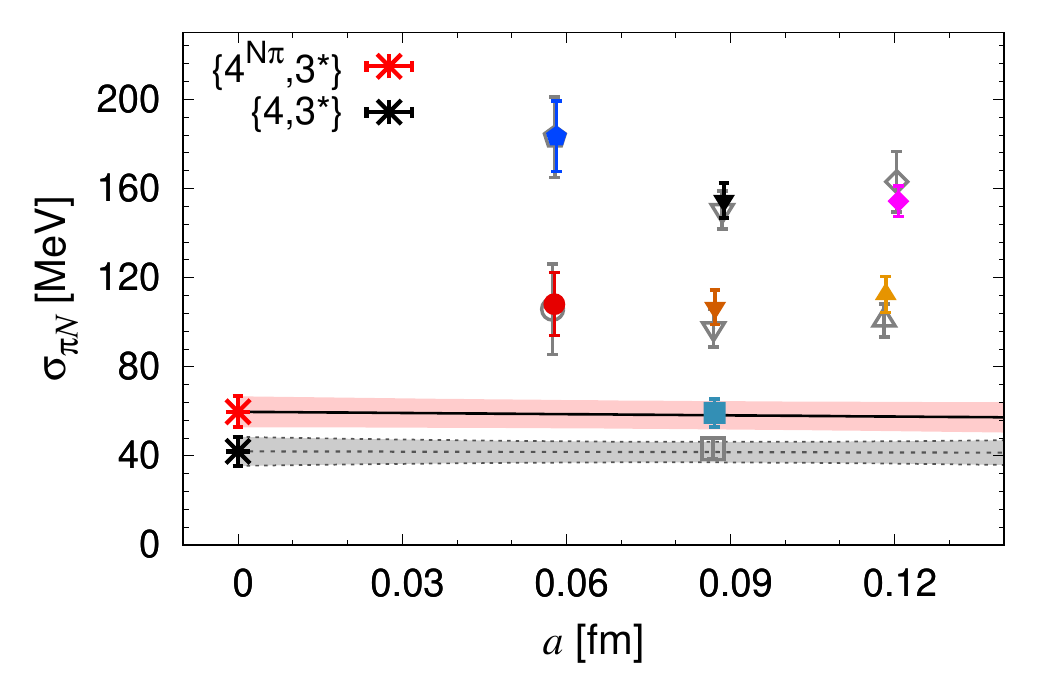}
  \includegraphics[width=0.24\linewidth]{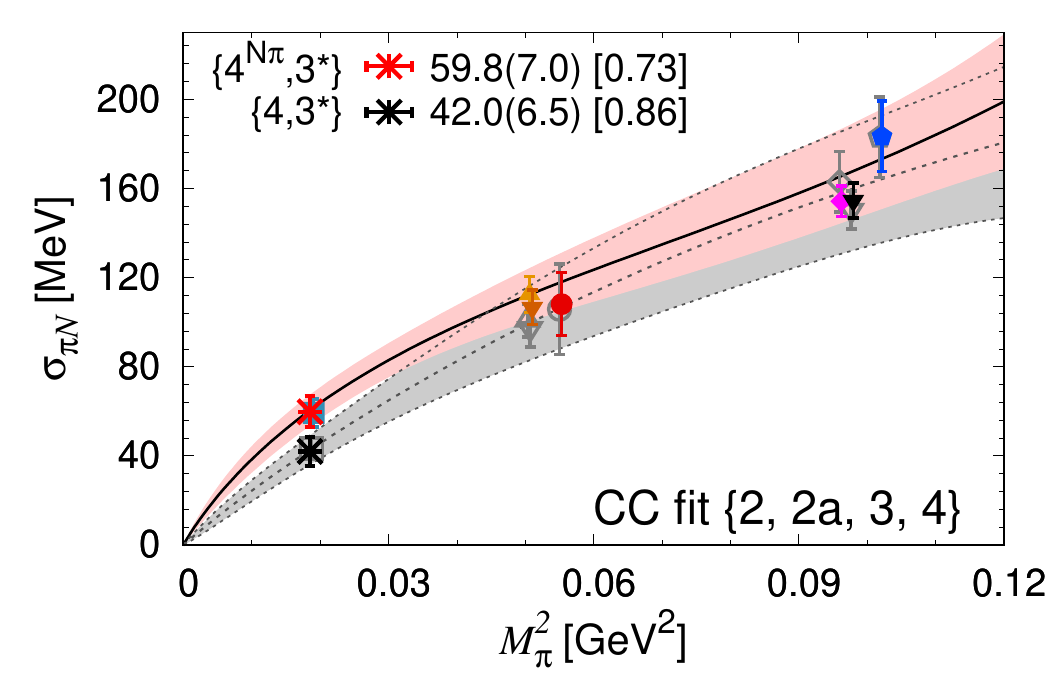}
  \includegraphics[width=0.24\linewidth]{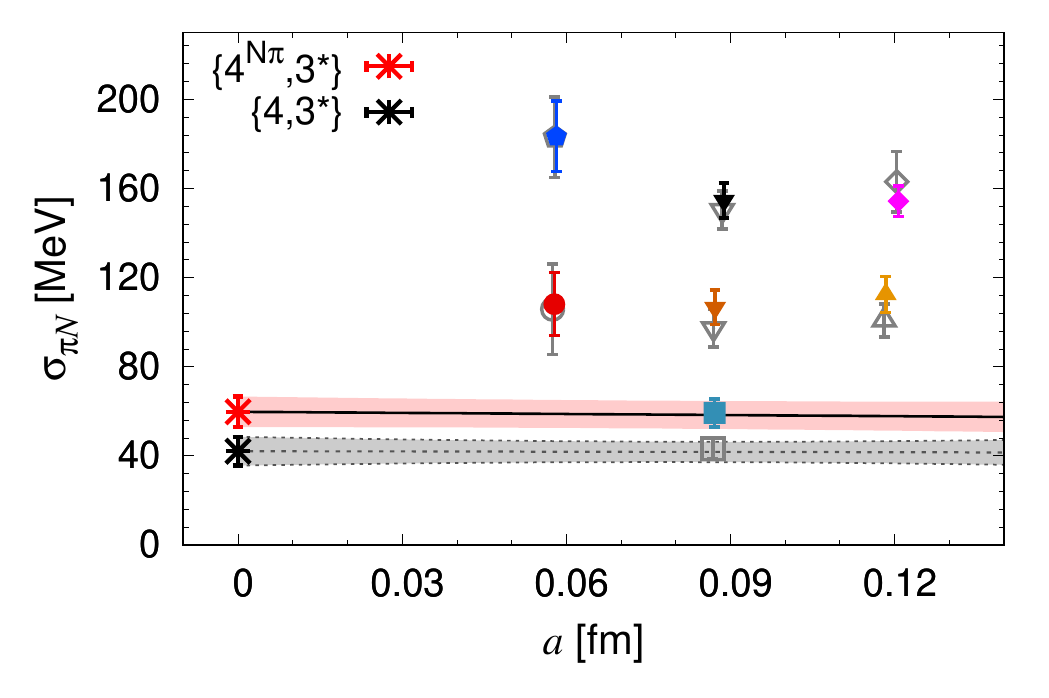}
  \includegraphics[width=0.24\linewidth]{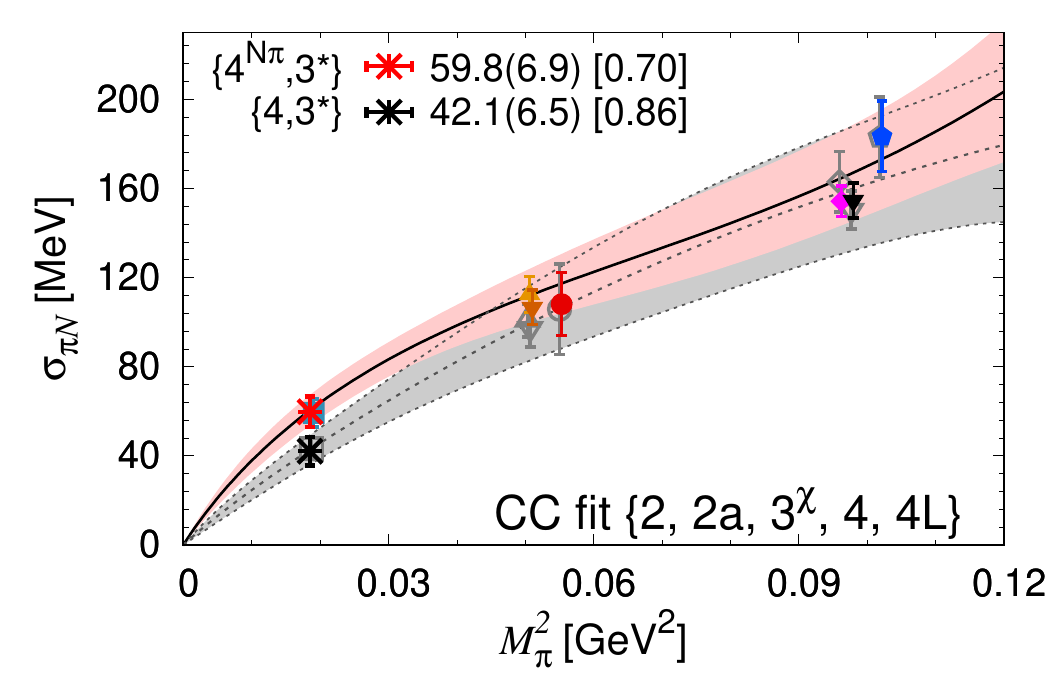}
  \vspace{-0.1in}
  \caption{CC fits to $\sigma_{N\pi}$ data with 2 different  chiral ansatz as explained in Fig.3 of Ref.~\cite{Gupta:2021ahb}.}
  \label{fig:CC_sigma}
\end{figure}

\textbf{The nucleon sigma term $\sigma_{\pi N}$ and the strangeness content $\sigma_s$:} 
The analysis of $\sigma_{\pi N}$ in
Ref.~\cite{Gupta:2021ahb} used the renormalization independent $\sigma_{\pi
  N}=m_l^\text{bare}g_S^{u+d,\text{bare}}$. The $\chi$PT analysis suggests that 4--5 $M_\pi$--dependent terms contribute significantly~\cite{Gupta:2021ahb}. With data at 
three values ($M_\pi \approx
135, 220, 310$~MeV), we have used different CC fit ansatz with 
some of the coefficients fixed to their values from $\chi$PT~\cite{Gupta:2021ahb}. Two such 
fits are shown in Fig.~\ref{fig:CC_sigma} to updated data with new  
analysis. \looseness-1
  
In this work, we use CC
extrapolated renormalized scalar charges $g_S^{u,d,s}$ and
renormalized quark masses $m_{l,s}$ ($N_f=2+1+1$ results from FLAG
2021 \cite{FLAG:2021npn}) to determine $\sigma_{\pi N}$ and
$\sigma_{s}$. These results are summarized in
Fig.~\ref{fig:flag_sigma} along with other lattice determinations.  The $N
\pi$ analysis gives $\sigma_{\pi N}|_{N\pi} \approx 60$~MeV,
consistent with phenomenology while the standard analysis value
$\sigma_{\pi  N}|_{\rm standard} \approx 40$~MeV is 
consistent with previous lattice estimates~\cite{Gupta:2021ahb}. For
$\sigma_s$, we use $g_S^s$ from the standard analysis, which gives $\sigma_s=38(9)$~MeV 
using $m_s=93.44(68)$~MeV from FLAG~\cite{FLAG:2021npn}.\looseness-1

\begin{table} 
  \centering
  \begin{tabular}{l | lll|ll}
    & \multicolumn{3}{c|}{This work (Preliminary)} & \multicolumn{2}{c}{PNDME'18}\\
    $q$ & $\phantom{-}g_A^q$ & $\phantom{-}g_T^q$ & $g_S^q$ & $\phantom{-}g_A^q$ \cite{Lin:2018obj} & $\phantom{-}g_T^q$ \cite{Gupta:2018lvp} \\
    \hline
    $u$ & $\phantom{-}0.79(3)(1)$ & $\phantom{-}0.75(3)(1)$ & 8.3(1.3) &  $\phantom{-}0.777(25)(30)$  &  $\phantom{-}0.784(28)(10)$\\
    $d$ & $-0.46(3)(3)$ & $-0.20(1)(1)$ & 8.5(9)   & $-0.438(18)(30)$ & $-0.204(11)(10)$\\
    $s$ & $-0.055(7)$   & $-0.0014(9)$  & 0.41(10)   & $-0.053(8)$      & $-0.00319(72)$\\
  \end{tabular}
  \caption{Updated preliminary results for the flavor diagonal charges compared to results published in~\cite{Gupta:2018lvp,Lin:2018obj}.}
  \label{tab:gAST}
\end{table}

\section{Conclusion} 
Significant progress has been made in calculating $\chi$PT predictions
and using them as guides in fits to remove ESC and to do the chiral extrapolation.
Unfortunately, excited-state fits to present data do not, in most cases, 
distinguish between standard and $N\pi$ analyses. To get percent level results for these charges with 
data driven methods to control ESC requires higher statistics at many more values of $a$ and
on physical pion mass ensembles.  

\pagebreak
\section{Acknowledgements}

We thank the MILC collaboration for providing the 2+1+1-flavor HISQ
lattices. The calculations used the Chroma software
suite~\cite{Edwards:2004sx}.  This research used resources at (i) NERSC, 
a DOE Office of Science facility supported under Contract No.\ DE-AC02-05CH11231; (ii)
the OLCF, a DOE Office of
Science User Facility supported under Contract DE-AC05-00OR22725, through 
ALCC awards LGT107 and INCITE awards PHY138 and HEP133; (iii) the USQCD
collaboration, which is funded by the Office of Science of the
U.S. DOE; and (iv) Institutional Computing at Los
Alamos National Laboratory. 
S.~Park acknowledges support from the U.S. Department of Energy Contract No. DE-AC05-06OR23177, under which Jefferson Science Associates, LLC, manages and operates Jefferson Lab. Also acknowledged is support from the Exascale Computing Project (17-SC-20-SC), a collaborative effort of the U.S. DOE Office of Science and the National Nuclear Security Administration.
T.~Bhattacharya and R.~Gupta were partly
supported by the U.S.\ DOE, Office of Science, HEP under Contract
No.\ DE-AC52-06NA25396. T.~Bhattacharya, R.~Gupta, S.~Mondal, S.~Park,
and B.~Yoon were partly supported by the LANL LDRD program, and
S.~Park by the Center for Nonlinear Studies.


\bibliographystyle{JHEP}
\let\oldbibitem\bibitem
\def\bibitem#1\emph#2,{\oldbibitem#1}
\let\oldthebibliography\thebibliography
\renewcommand\thebibliography[1]{\oldthebibliography{#1}%
                                 \itemsep0pt\parskip0pt\relax}
\bibliography{ref}

\end{document}